\documentclass[twocolumn]{revtex4-1}
\usepackage[utf8]{inputenc}
\usepackage{graphicx}
\usepackage{braket}
\usepackage{amsmath}
\usepackage{hyperref}

\begin{document}
\title{Stability and superconductivity of lanthanum and yttrium decahydrides}

\author{Alice M. Shipley}
\email{ams277@cam.ac.uk}
\affiliation
{
    Theory of Condensed Matter Group,
    Cavendish Laboratory,
    J.~J.~Thomson Avenue,
    Cambridge CB3 0HE,
    United Kingdom
}

\author{Michael J. Hutcheon}
\email{mjh261@cam.ac.uk}
\affiliation
{
    Theory of Condensed Matter Group,
    Cavendish Laboratory,
    J.~J.~Thomson Avenue,
    Cambridge CB3 0HE,
    United Kingdom
}

\author{Mark S. Johnson}
\affiliation
{
    Theory of Condensed Matter Group,
    Cavendish Laboratory,
    J.~J.~Thomson Avenue,
    Cambridge CB3 0HE,
    United Kingdom
}

\author{Chris J. Pickard}
\affiliation
{
    Department of Materials Science and Metallurgy,
    27 Charles Babbage Rd,
    Cambridge CB3 0FS,
    United Kingdom
}
\affiliation
{
    Advanced Institute for Materials Research, 
    Tohoku University, 2-1-1 Katahira, 
    Aoba, Sendai, 980-8577, 
    Japan
}

\author{Richard J. Needs}
\affiliation
{
    Theory of Condensed Matter Group,
    Cavendish Laboratory,
    J.~J.~Thomson Avenue,
    Cambridge CB3 0HE,
    United Kingdom
}


\begin{abstract}
Rare-earth hydrides can exhibit high-temperature superconductivity under high pressure. Here, we apply a crystal structure prediction method to the current record-holding $T_c$ material, LaH$_{10}$. We find a pressure-induced phase transition from the experimentally observed cubic phase to a hexagonal phase at around 420\ GPa. This new phase is metastable down to low pressures and could explain experimental observations of hcp impurities in fcc samples. We go on to find that YH$_{10}$ adopts similar structures and discuss the sensitivity of superconductivity calculations to certain computational parameters.
\end{abstract}

\maketitle

\section{Introduction}

Hydrogen was predicted to be a room-temperature superconductor at very high pressure in 1968 \cite{ashcroft1968}, but the pressures required to metallise hydrogen are difficult to obtain \cite{mcmahon2012,mao1989,eremets2011,dalladay2016,dias2017,loubeyre2020}. Hydrides have been suggested to have lower metallisation pressures than pure hydrogen due to \textit{chemical pre-compression} \cite{ashcroft2004} and therefore might become superconducting at more readily accessible pressures. This idea has motivated a surge of research examining potential superconductivity in high-pressure hydrides, with several reviews summarising recent developments \cite{duan2017review, zurek2019, flores2019, boeri2019, oganov2019, needs2016, pickard2019}.

Theoretical studies of ScH$_3$, LaH$_3$ \cite{durajski2014}, YH$_3$, YH$_4$ and YH$_6$ \cite{kim2009, li2015, heil2019} identified hydrides of rare-earth elements as potential high-temperature superconductors. First-principles structure searching studies of rare-earth hydrides have reported structures with high hydrogen content adopting cage-like structures \cite{liu2017, peng2017}. Of particular note, a $T_c$ of 264-286\ K was calculated for $Fm\bar{3}m$ LaH$_{10}$ at 210\ GPa \cite{liu2017}, while the analogous YH$_{10}$ structure was calculated to have $T_c$ = 305-326\ K at 250\ GPa. Slight distortions of the cubic LaH$_{10}$ phase were found to lead to $C2/m$ and $R\bar{3}m$ structures at lower pressures \cite{geballe2018,liu2018}, though Ref.\ \cite{errea2019} showed that quantum effects render $Fm\bar{3}m$ as the true ground state. These predictions were followed by experimental measurement of critical temperatures reaching 260\ K in LaH$_{10}$ at 170-200\ GPa \cite{somayazulu2019, drozdov2018v2}. The high-$T_c$ phase was determined to be a structure with an fcc arrangement of La atoms, lending support to theoretical predictions.

In addition to the aforementioned studies, others have focused on heavier rare-earth hydrides, exploring the synthesis and superconducting properties of cerium \cite{salke2019, li2019}, praseodymium \cite{zhou2019} and neodymium \cite{zhou2019neodymium} hydrides. Here, within the framework of density functional theory (DFT) \cite{hohenberg1964,kohn1965}, we revisit LaH$_{10}$ and YH$_{10}$ using crystal structure prediction methods. We find a phase transition to a new hexagonal phase in LaH$_{10}$ at high pressures, with the metastability of this phase at low pressures offering an explanation for the experimental observation of hcp impurities in fcc samples \cite{drozdov2018v2}. We go on to predict the phases and corresponding critical temperatures that may be observed in YH$_{10}$.

\section{Theory and methodology}

\subsection{Phonons and superconductivity}
The Hamiltonian of a coupled electron-phonon system \cite{DFT_ELEC_PHONON} can be written as 
\begin{equation}
\label{eq:electron_phonon_hamiltonian}
\begin{aligned}
    H &= \underbrace{\sum_{kn} \epsilon_{nk} c_{nk}^\dagger c_{nk}}_{\text{electronic dispersion}} + \underbrace{\sum_{q\nu} \omega_{q\nu} \left(a_{q\nu}^\dagger a_{q\nu} + \frac{1}{2}\right)}_{\text{phonon dispersion}} +\\
    &\underbrace{\frac{1}{\sqrt{N_p}} \sum_{kqmn\nu} g_{mn\nu}(k,q)c_{m,k+q}^\dagger c_{nk} \left( a_{q\nu} + a_{-q\nu}^\dagger\right).}_{\text{electron-phonon coupling}}
\end{aligned}
\end{equation}
In this work, we calculate the electronic Kohn-Sham eigenvalues $\epsilon_{nk}$, phonon frequencies $\omega_{q,\nu}$, and electron-phonon coupling constants $g_{mn\nu}(k,q)$ appearing in $H$ from first-principles using the \textsc{quantum espresso} DFT code \cite{QE-2009,QE-2017}, which we optimised for this work \footnote{Our optimisations have been submitted to the \textsc{quantum espresso} project.}. The Hamiltonian in Eq.\ \ref{eq:electron_phonon_hamiltonian} can be treated within Migdal-Eliashberg theory \cite{eliashberg1960}, allowing us to define the electron-boson spectral function
\begin{equation}
\begin{aligned}
\label{eq:eliashberg_function}
\alpha^2F(\omega) &= \frac{1}{N(\epsilon_F)} \sum_{mnq\nu} \delta(\omega - \omega_{q\nu})\sum_k |g_{mn\nu}(k,q)|^2 
\\ &\times\delta(\epsilon_{m,k+q}-\epsilon_F) \delta(\epsilon_{n,k}-\epsilon_F).
\end{aligned}
\end{equation}
From $\alpha^2F$ we extract the superconducting critical temperature by solution of the Eliashberg equations using the \textsc{elk} code \cite{elk_code}. From the quantities appearing in $H$ we may also construct the electronic and vibrational densities of states, from which we can derive the Gibbs free energy as a function of temperature. We do this at a range of pressures, allowing us to construct pressure-temperature phase diagrams.

\begin{figure}
    \centering
    \includegraphics[width=\columnwidth]{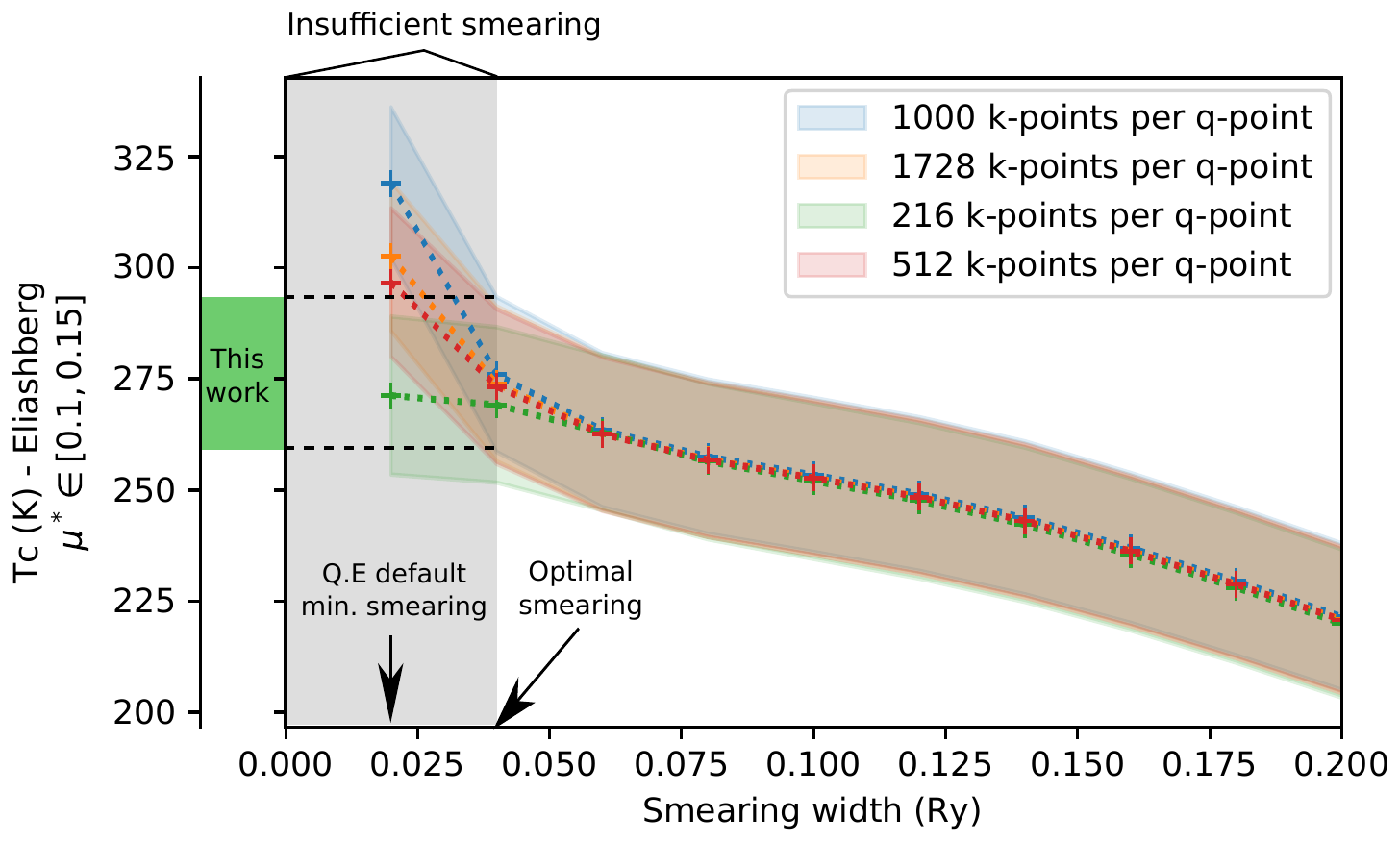}
    \caption{The dependence of $T_c$ on the double-delta smearing width, $\sigma$, for $Fm\bar{3}m$-YH$_{10}$ at 350\ GPa. The region of insufficient smearing is shown, along with our choice of $\sigma$ for this structure and pressure. The smallest value used in an electron-phonon calculation with default \textsc{quantum espresso} settings is shown.}
    \label{fig:yh10_smearing}
\end{figure}

\begin{figure}
    \centering
    \includegraphics[width=\columnwidth]{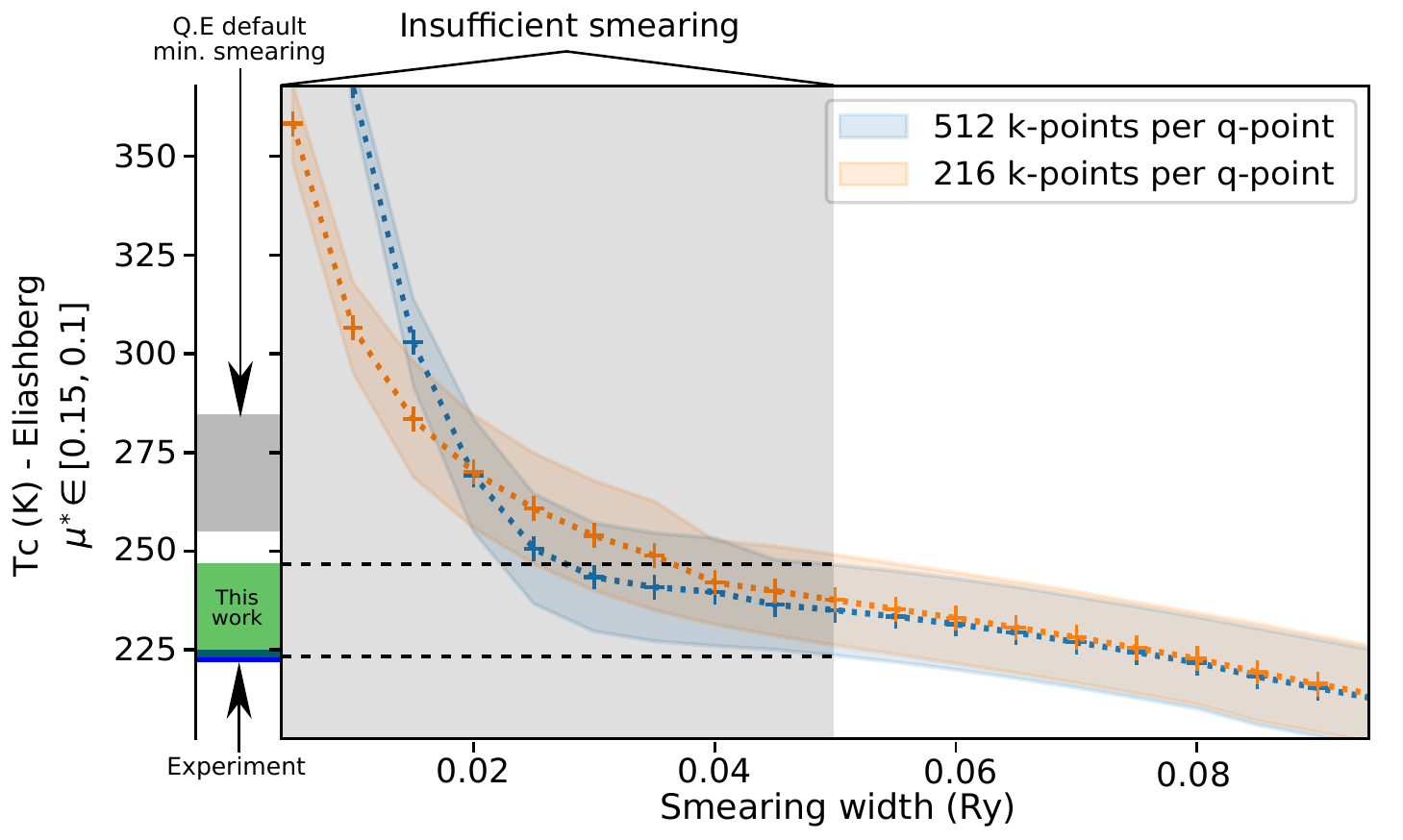}
    \caption{The dependence of $T_c$ on the double-delta smearing width, $\sigma$, for $Im\bar{3}m$-YH$_6$ at 160\ GPa. A recent experimental measurement at 166\ GPa, falling just within our calculated $T_c$ range, is also shown \cite{troyan2019}. We note that Refs.\ \cite{troyan2019,kong2019} highlighted that previous calculated $T_c$ values were considerably higher than their experimental observations and that the results of Ref.\ \cite{heil2019}, which used accurate Wannier interpolation techniques, are in agreement with ours.}
    \label{fig:yh6_smearing}
\end{figure}

To evaluate the double-delta sum in Eq.\ \ref{eq:eliashberg_function} for finite $\mathbf{k}$- and $\mathbf{q}$-point grids, we follow the method detailed in Appendix A of Ref.\ \cite{wierzbowska2005} and smear the delta functions with finite-width Gaussians. In order to best approximate the delta functions, the smallest sensible smearing should be used. However, the smearing must be large enough to accommodate the finite $\mathbf{k}$-point grids used. We identify the optimal choice of smearing from discrepancies in the results between different $\mathbf{k}$-point grids \cite{wierzbowska2005}, as can be seen in Figs.\ \ref{fig:yh10_smearing} and \ref{fig:yh6_smearing}.

Our electron-phonon calculations were carried out using the Perdew-Burke-Ernzerhof (PBE) generalised gradient approximation \cite{pbe1996} and ultrasoft pseudopotentials, validated against the all-electron WIEN2k code \cite{wien2k_code,supplement}. Well-converged $\mathbf{k}$-point grids with a spacing of at most $2\pi\times$0.015\ \r{A}$^{-1}$ and an 820\ eV plane wave cut-off were used \cite{supplement}. The $\mathbf{q}$-point grids used were typically 8 times smaller than the $\mathbf{k}$-point grids and were Fourier interpolated to 10 times their original size. For the cubic systems studied, this corresponds to $\geq 24 \times 24 \times 24$ $\mathbf{k}$-point grids and a $3 \times 3 \times 3$ $\mathbf{q}$-point grid Fourier-interpolated to $30 \times 30 \times 30$.

\subsection{Structure searching}
Our structure searching calculations were performed using \textit{ab initio} random structure searching (AIRSS) \cite{pickard2011, needs2016} and \textsc{castep} \cite{castep2005}. The PBE functional, \textsc{castep} QC5 pseudopotentials, a 400\ eV plane wave cut-off and a $\mathbf{k}$-point spacing of $2\pi\times$0.05\ \r{A}$^{-1}$ were used in these searches unless otherwise stated. The \textsc{c2x} software \cite{c2x} was used for converting between \textsc{castep} and \textsc{quantum espresso} file formats, and also for reporting the space groups of structures at various tolerances.

\section{Results and discussion}
In the following sections, we report results in terms of phonon-corrected pressures, obtained by fitting the Birch-Murnaghan equation of state \cite{birch_murnaghan} to our data. Where static DFT pressures are reported instead, they are labelled as $P_{DFT}$ - this second set of pressures facilitates comparison with previous calculations as they are simply an input parameter to the DFT geometry optimisation.

\subsection{\texorpdfstring{LaH$_{10}$}{LaH10}}

\begin{figure}
    \centering
    \includegraphics[width=\columnwidth]{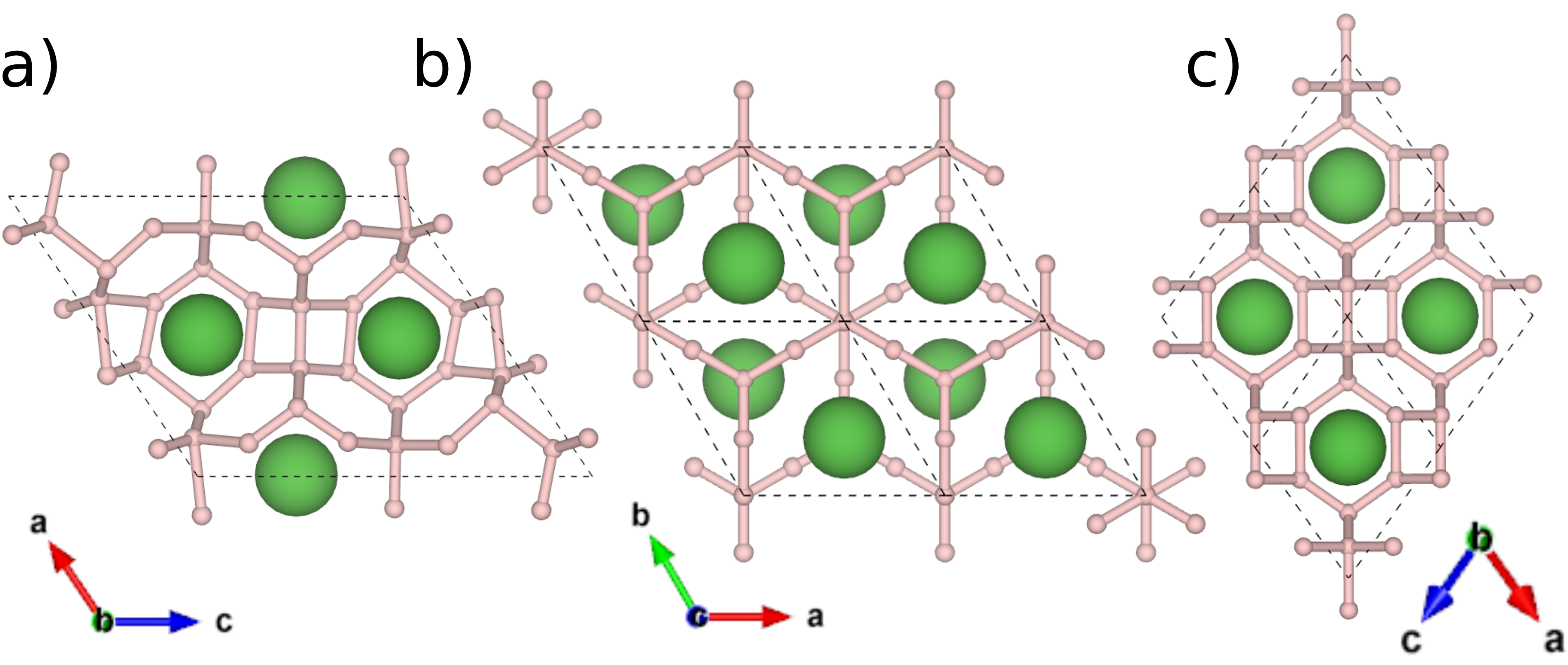}
    \caption{Structures of LaH$_{10}$. (a) 2 formula unit/cell $C2/m$, (b) 2 formula unit/cell $P6_3/mmc$, (c) 1 formula unit/cell $Fm\bar{3}m$. The $R\bar{3}m$ structure is not shown as it is visually indistinguishable from the $Fm\bar{3}m$ structure at the pressures of interest.}
    \label{fig:lah10_structures}
    \vspace{0.25cm}
    \includegraphics[width=0.9\columnwidth]{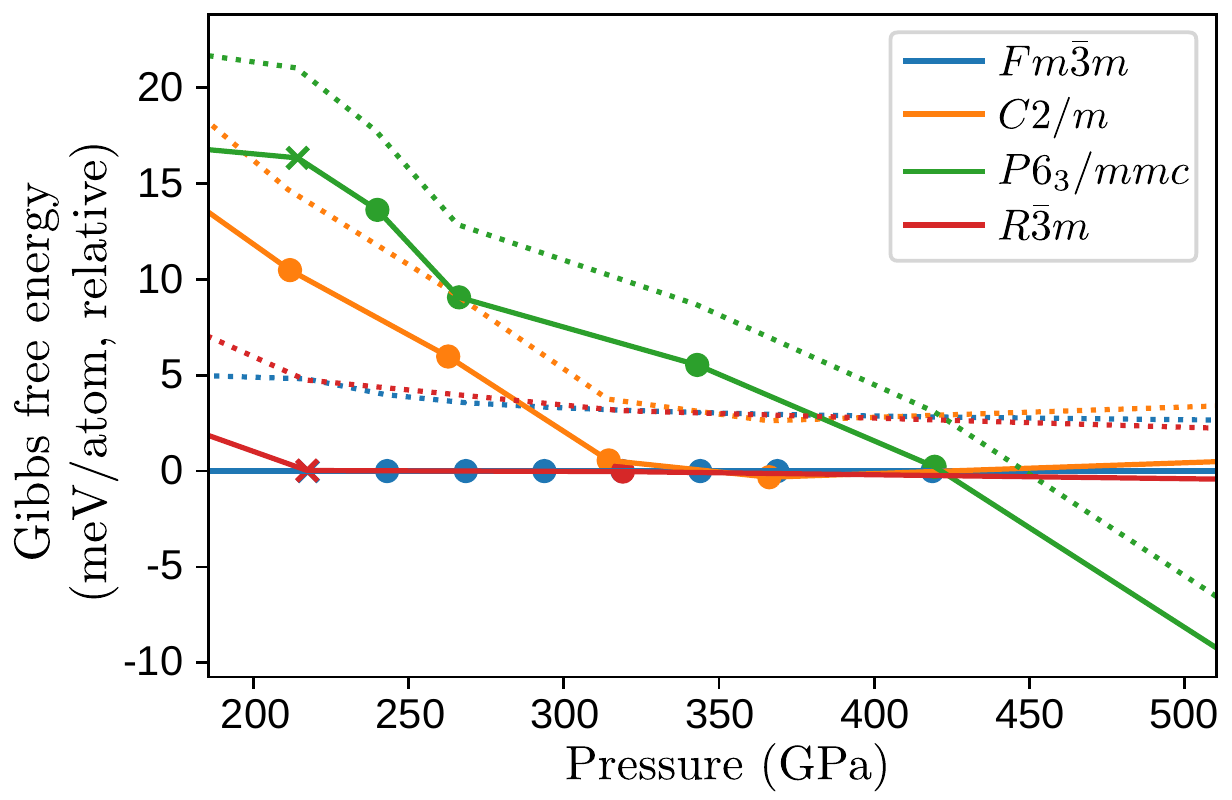}
    \caption{The Gibbs free energy as a function of pressure for energetically competitive phases of LaH$_{10}$, plotted relative to a third-order Birch-Murnaghan fit of the $Fm\bar{3}m$ data. Crosses represent calculations with unstable phonon modes - these points are not included in the Gibbs free energy fit. Solid lines are at 300\ K, dashed lines are at 0\ K.}
    \label{fig:lah10_thermo}
    \vspace{0.25cm}
    \includegraphics[width=0.9\columnwidth]{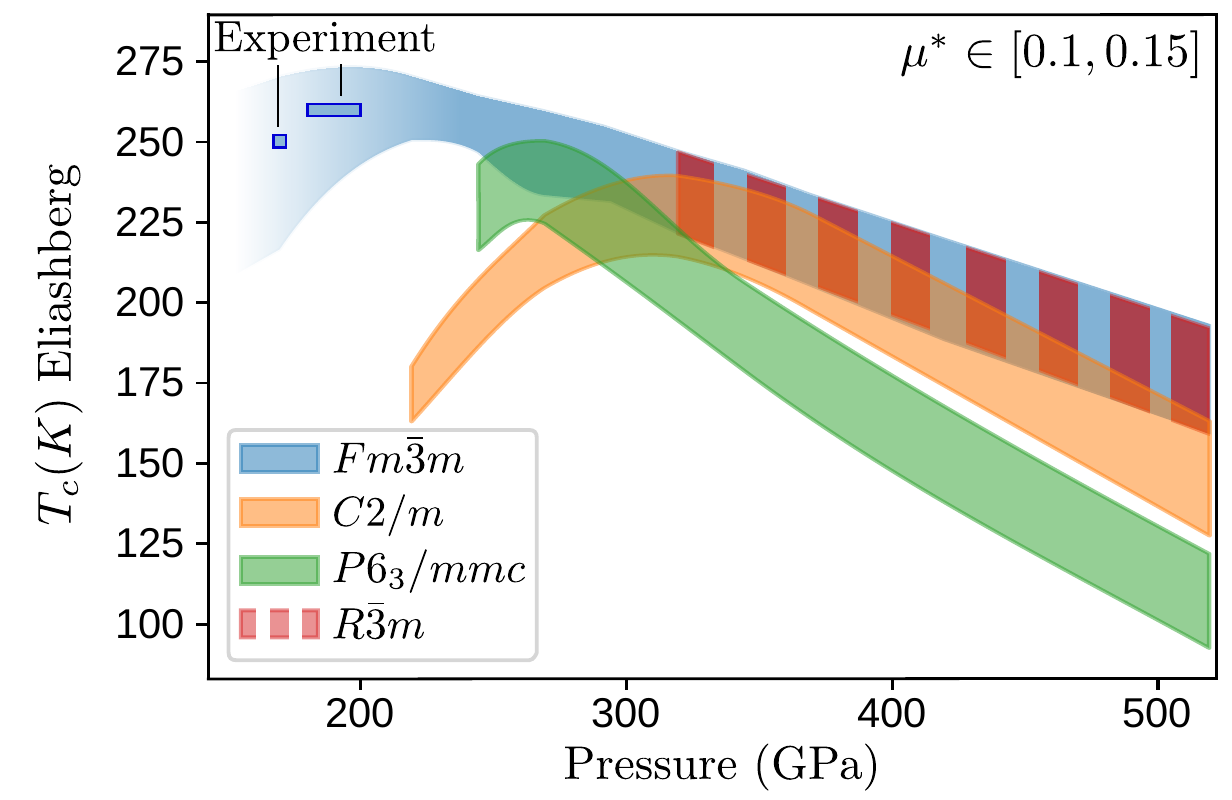}
    \caption{Calculated $T_c(P)$ for dynamically stable phases of LaH$_{10}$ from direct solution of the Eliashberg equations. The width of the lines arises from our treatment of the Morel-Anderson pseudopotential, $\mu^*$, \cite{mu_star} as an empirical parameter with values between 0.1 and 0.15. The $Fm\bar{3}m$ result has been extended into the region where it is dynamically unstable (shaded according to unstable fraction of the phonon density of states) in order to facilitate comparison with the experimental results of Refs.\ \cite{drozdov2018v2, somayazulu2019}. This extension was achieved by removing the contribution of unstable phonon modes, in their entirety, to the Eliashberg function while maintaining its normalisation.}
    \label{fig:lah10_tc}
\end{figure}

\begin{figure}
    \centering
    \includegraphics[width=0.9\columnwidth]{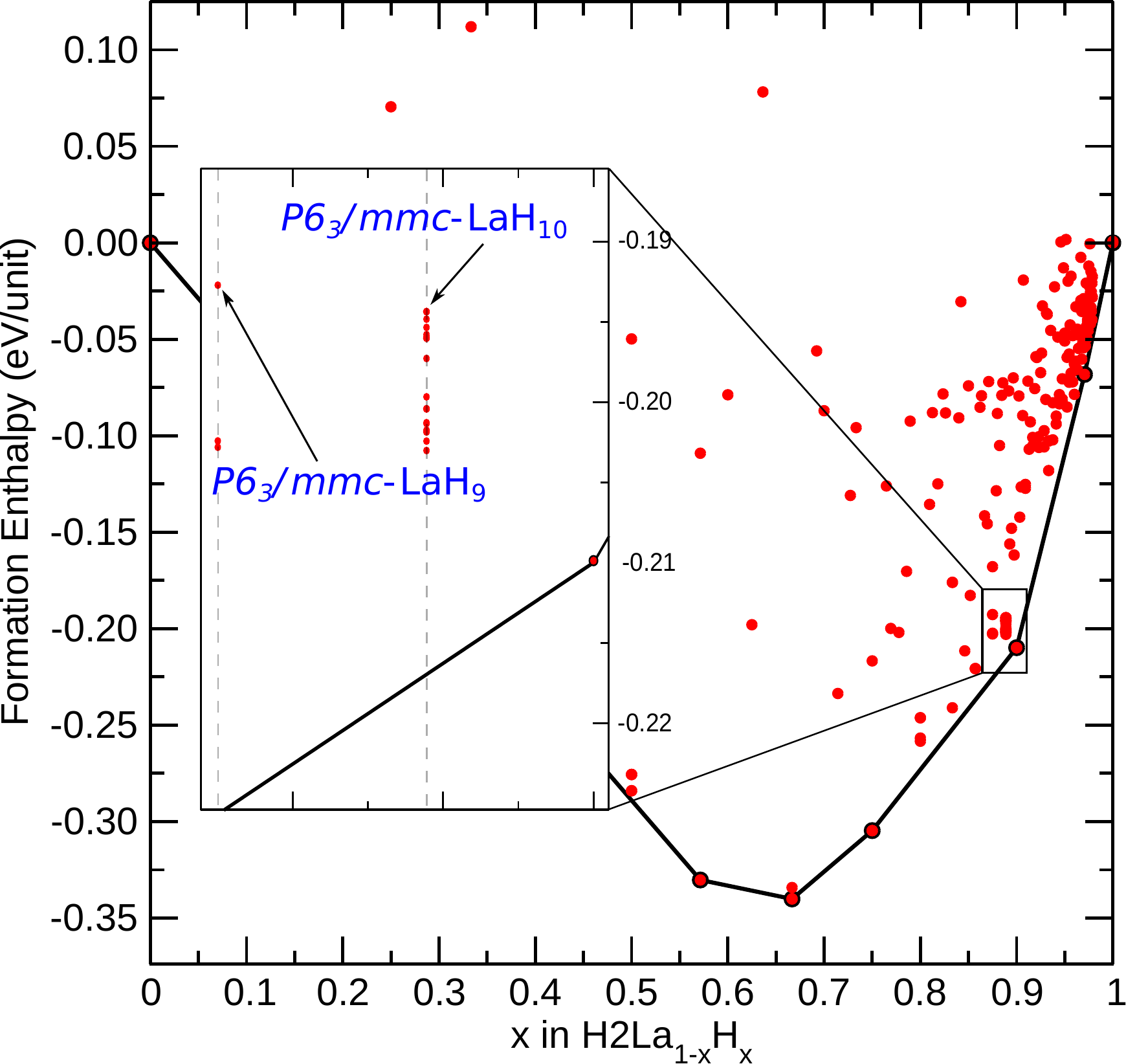}
    \caption{A convex hull for the La-H system at 150\ GPa, accurately calculated using \textsc{castep}, $\mathbf{k}$-point spacing of $2\pi\times$0.03\ \r{A}$^{-1}$ and a 700\ eV plane-wave cut-off. The on-the-fly pseudopotential strings used are provided in the supplementary material \cite{supplement}; the inclusion of a fraction of a 4f electron in the generation of the La pseudopotential was found to be crucial. A pseudopotential without this addition was unable to reproduce the all-electron $Fm\bar{3}m$-LaH$_{10}$ PV curve and led to a qualitatively different convex hull \cite{supplement}. In agreement with Ref.\ \cite{kruglov2020}, we find that LaH$_9$ is not on the hull at this pressure. However, we also find that LaH$_{16}$ does not lie on the hull at 150\ GPa, despite finding the $P6/mmm$-LaH$_{16}$ structure studied in that work.}
    \label{fig:lah_hull150}
\end{figure}

Low enthalpy candidates found by AIRSS for LaH$_{10}$ include the space groups $Fm\bar{3}m$, $R\bar{3}m$, and a 2-formula-unit $C2/m$, which had been identified previously. The searches also revealed a new structure of $P6_3/mmc$ symmetry. These structures are shown in Fig.\ \ref{fig:lah10_structures}. A $C2/m$ structure with 3 formula units per unit cell was also found to be energetically competitive, but was not considered further in this work as it behaves similarly to the 2-formula unit phase over the pressure range of interest. We also found several previously unreported structures at low pressures with space groups $Pnnm$, $C2$ and $P2_12_12_1$ and unit cells containing 2, 3 and 4 formula units, respectively. These are the lowest enthalpy structures in the low pressure region \cite{supplement}. However, we note that these structures are distortions of the high-symmetry $Fm\bar{3}m$ structure and, similarly to the case of $R\bar{3}m$ noted in Ref.\ \cite{errea2019}, it is possible that anharmonic effects may remove them from the potential energy surface. In addition to this, the low symmetry and large unit cells of these structures make converged phonon calculations prohibitively expensive; they are therefore not considered further in this work.


The calculated LaH$_{10}$ phase behaviour is shown in Fig.\ \ref{fig:lah10_thermo} and the corresponding critical temperatures are shown in Fig.\ \ref{fig:lah10_tc}. Our calculations for the $Fm\bar{3}m$ phase include unstable phonon modes for $P_{DFT}\leq$210\ GPa. In the harmonic picture, explicitly taking into account this dynamical instability leads to a window of stability for the $C2/m$ phase \cite{supplement}, which is in agreement with previous calculations \cite{geballe2018,liu2018}. However, we note that under the assumption that the unstable modes can be neglected in the calculation of the Gibbs free energy, we obtain the same behaviour as the anharmonic calculations of Ref.\ \cite{errea2019}, i.e., $Fm\bar{3}m$ is the only phase with a predicted region of stability at lower pressures. With increasing pressure, as noted in previous theoretical work \cite{errea2019}, the $R\bar{3}m$ structure approaches $Fm\bar{3}m$ symmetry. We therefore expect that these phases will not be distinguishable at high pressures.

At 300\ K, the $P6_3/mmc$ structure becomes thermodynamically favourable at pressures above $\sim$420\ GPa. More importantly, this hexagonal phase is also metastable at low pressures, lying within 20\ meV/atom of the cubic phase down to 150\ GPa, and therefore provides an explanation for the experimental observation of hcp impurities in fcc-LaH$_{10}$ samples at 170\ GPa in Ref.\ \cite{drozdov2018v2}.

A low-energy hexagonal LaH$_9$ structure predicted previously in similar pressure regions \cite{kruglov2020} could offer an alternative explanation for the observation of these impurities. However, the authors of Ref.\ \cite{drozdov2018v2} determined that the two kinds of hcp impurities in their fcc-LaH$_{10}$ samples possessed LaH$_{10}$ stoichiometry. We also calculated a high-quality La-H convex hull at 150\ GPa using AIRSS \cite{pickard2011} and qhull \cite{barber1996} (see Fig.\ \ref{fig:lah_hull150}). It shows that the $P6_3/mmc$-LaH$_{10}$ structure predicted in this work lies closer to the hull than the $P6_3/mmc$-LaH$_9$ structure of Ref.\ \cite{kruglov2020}. It is therefore likely that the hcp impurities originate from our new $P6_3/mmc$-LaH$_{10}$ phase.

To facilitate comparison with experiment, we have calculated powder X-ray diffraction patterns for cubic and hexagonal LaH$_{10}$ and hexagonal LaH$_9$ at 150\ GPa - we supply these, alongside the calculated c/a ratios and volumes in the supplementary material \cite{supplement}.

We calculate $T_c$ = 232-259\ K for $Fm\bar{3}m$-LaH$_{10}$ at 269\ GPa ($P_{DFT}$=250\ GPa), which is lower than the previous theoretical result of $T_c$ = 257-274\ K \cite{liu2017}. However, we observe an increase in $T_c$ on reduction of the double-delta smearing parameter to below our calculated optimal value \cite{supplement}, potentially explaining this discrepancy. Careful choice of smearing has previously been noted as important in other hydride systems \cite{heil2018}. We also note a previous calculation of $T_c$ for this structure at 200\ GPa \cite{peng2017}, however, in agreement with other calculations \cite{liu2018, geballe2018} we find $Fm\bar{3}m$ to be dynamically unstable at this pressure. This dynamical instability means we cannot directly compare with experiment, which found $T_c$ = 250\ K at around 170\ GPa \cite{drozdov2018v2} and $T_c$ = 260\ K at 180-200\ GPa \cite{somayazulu2019}. However, ignoring the contribution of the unstable phonon modes to the Eliashberg function at pressures $\leq$ 210\ GPa allows for a rough estimation of $T_c$ in these regions; this is depicted as the faded-out section in Fig.\ \ref{fig:lah10_tc} and the results obtained are in agreement with experimental results. For the $C2/m$ phase, using an optimal value of smearing we calculate $T_c$ = 205-225\ K at 262\ GPa ($P_{DFT}$=250\ GPa), compared to $T_c$ = 229-245\ K in Ref.\ \cite{liu2018}.

\subsection{\texorpdfstring{YH$_{10}$}{YH10}}

\begin{figure}
    \centering
    \includegraphics[width=\columnwidth]{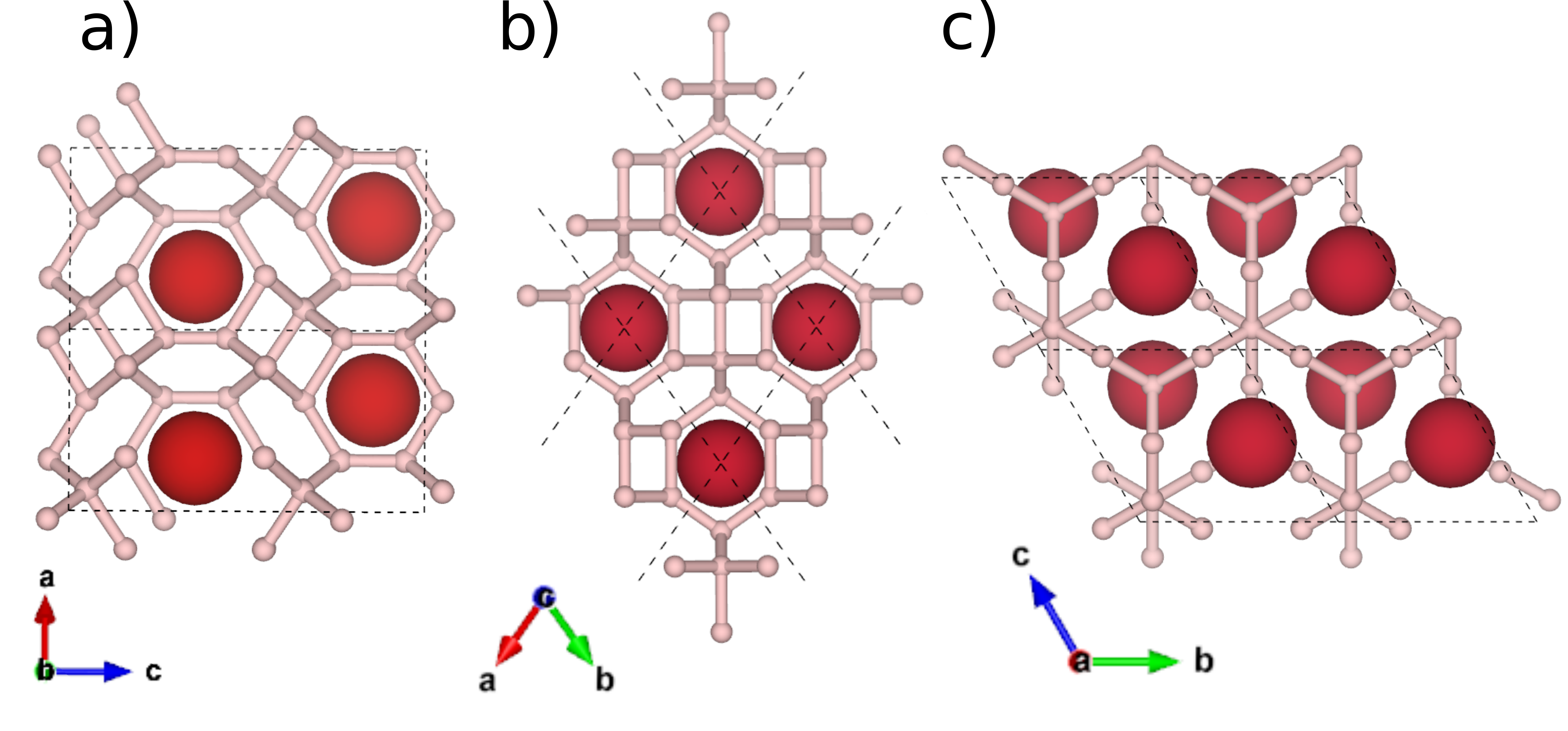}
    \caption{Structures of YH$_{10}$. (a) 2 formula unit/cell $Cmcm$, (b) 1 formula unit/cell $Fm\bar{3}m$, (c) 2 formula unit/cell $P6_3/mmc$. The $R\bar{3}m$ structure is, again, not shown.}
    \label{fig:yh10_structures}
    \includegraphics[width=0.9\columnwidth]{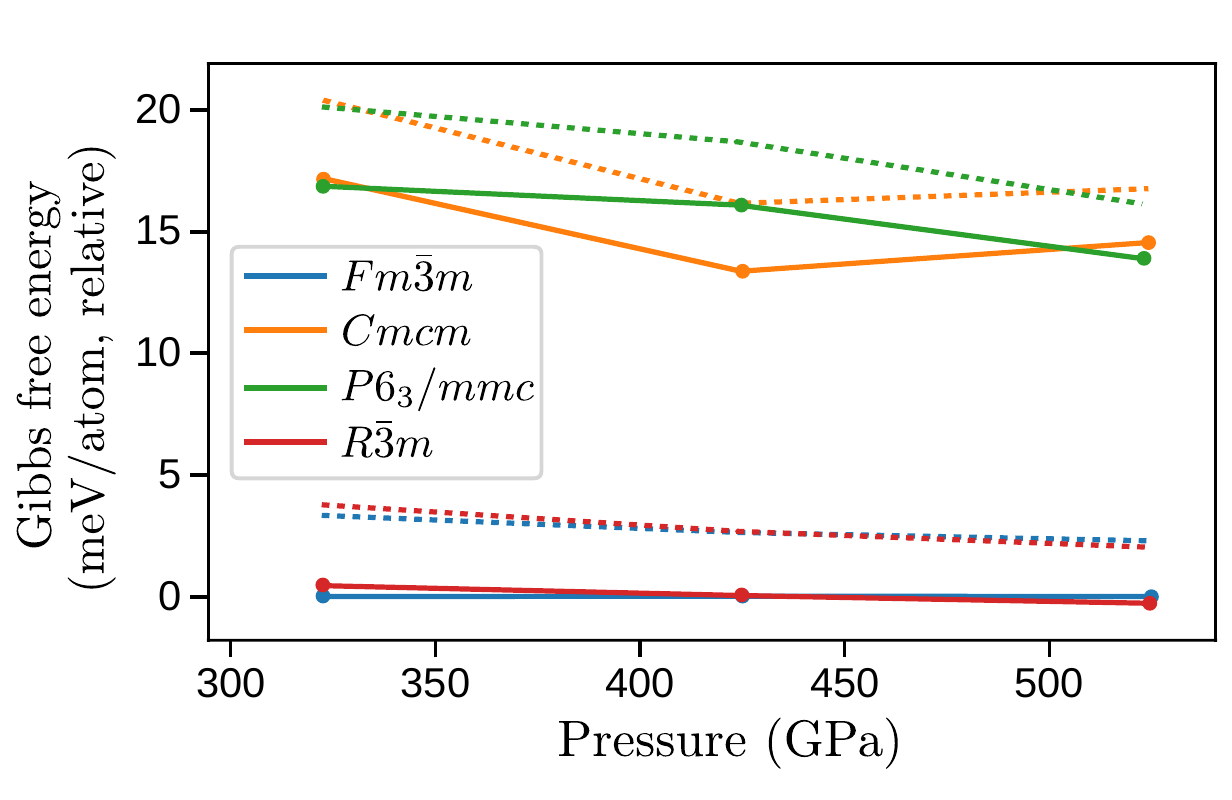}
    \caption{The Gibbs free energy as a function of pressure for energetically competitive phases of YH$_{10}$, plotted and interpolated relative to a third-order Birch-Murnaghan fit of the $Fm\bar{3}m$ data. Solid lines are at 300\ K, dashed lines are at 0\ K.}
    \label{fig:y10_thermo}
    \vspace{0.25cm}
    \includegraphics[width=0.9\columnwidth]{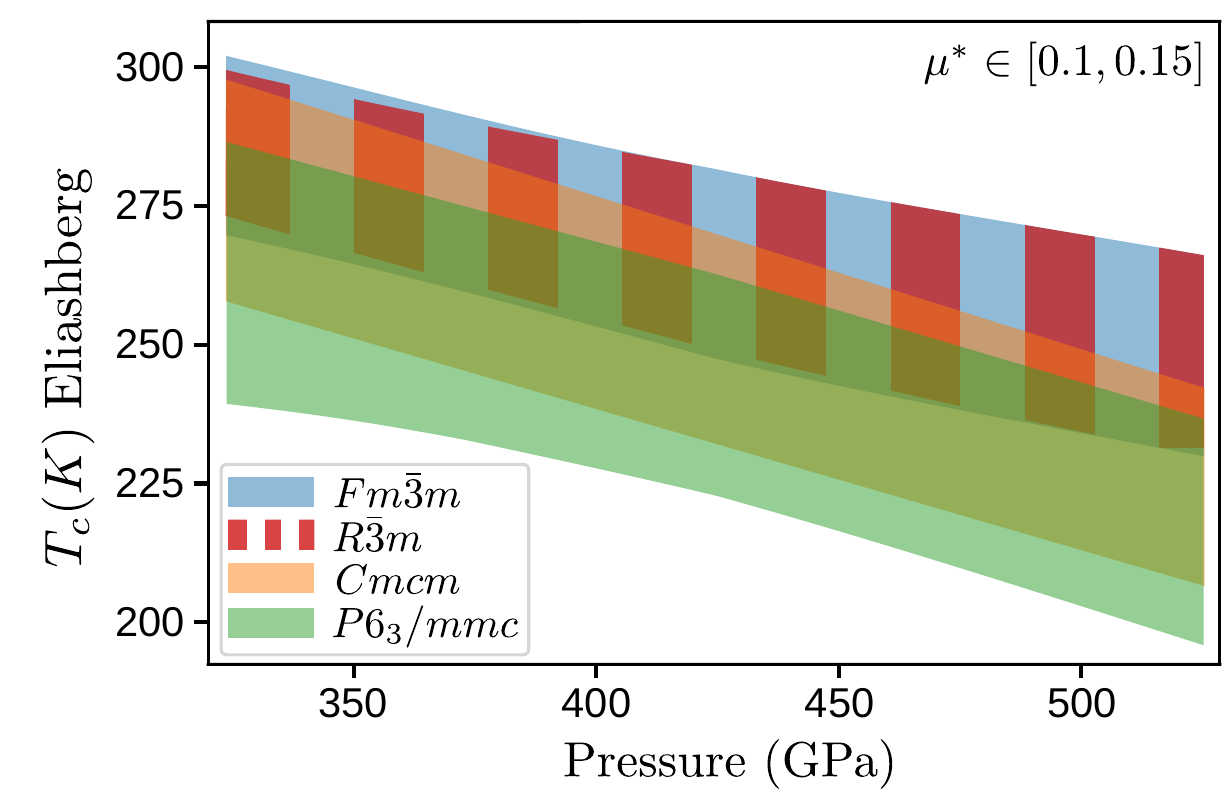}
    \caption{Calculated $T_c(P)$ for dynamically stable phases of YH$_{10}$ from direct solution of the Eliashberg equations. $\mu^*$ is, again, taken to have a value between 0.1 and 0.15.}
    \label{fig:yh10_tc}
\end{figure}

Low-enthalpy candidates for YH$_{10}$ found using AIRSS include $Fm\bar{3}m$, which had been identified previously, a slight distortion of this phase, $R\bar{3}m$, and structures of $P6_3/mmc$ and $Cmcm$ symmetry. These structures are shown in Fig.\ \ref{fig:yh10_structures}. The calculated YH$_{10}$ phase behaviour is shown in Fig.\ \ref{fig:y10_thermo} and the corresponding critical temperatures are shown in Fig.\ \ref{fig:yh10_tc}. We do not predict any phase transitions within the predicted range of stability of the YH$_{10}$ stoichiometry \cite{peng2017}. However, the difference in Gibbs free energy between the $Fm\bar{3}m$ and $R\bar{3}m$ phases is exceedingly small (see Fig.\ \ref{fig:y10_thermo}), reflecting their structural similarity.


Previous calculations for $Fm\bar{3}m$ found $T_c$ = 305-326\ K at 250\ GPa \cite{liu2017} and $T_c$ = 303\ K at 400\ GPa \cite{peng2017}. Here, we calculate $T_c$ = 270-302\ K at 324\ GPa ($P_{DFT}$=300\ GPa) and $T_c$ = 250-280\ K at 425\ GPa ($P_{DFT}$=400\ GPa). Our more conservative $T_c$ results can again be explained by considering the smearing parameter used to approximate the double-delta integral in Eq.\ \ref{eq:eliashberg_function}. We were able to reproduce the results of previous calculations by using the minimum default smearing employed in \textsc{quantum espresso}, which in this case overestimates $T_c$ by $\sim$30\ K (see Fig.\ \ref{fig:yh10_smearing}) compared to optimal smearing. We note that our results agree with those obtained using Wannier interpolation techniques \cite{heil2019}. Using the same method to calculate an optimal smearing also provides results in agreement with recent experimental measurements for $Im\bar{3}m$-YH$_6$ \cite{troyan2019}, as shown in Fig.\ \ref{fig:yh6_smearing}.

\section{Conclusions}
We have identified a new hexagonal phase of LaH$_{10}$ with $P6_3/mmc$ symmetry. Our calculations show a pressure-induced phase transition into this new phase from the cubic phase believed to be observed in experiment \cite{somayazulu2019, drozdov2018v2}. The overall phase behaviour predicted within the harmonic picture is $C2/m \rightarrow Fm\bar{3}m \rightarrow P6_3/mmc$ with all three of these phases predicted to be high-$T_c$ superconductors. Making the assumption that unstable modes can be neglected, however, gives the same picture as the anharmonic calculations of Ref.\ \cite{errea2019} where $Fm\bar{3}m$ is the true ground state at lower pressures. The new hexagonal phase predicted here offers a direct explanation for the observation of hcp impurities in recent experiments \cite{drozdov2018v2}.
                                             
We found that YH$_{10}$ adopts very similar structures to LaH$_{10}$, with one of $P6_3/mmc$ symmetry again amongst the most energetically competitive candidates. Over the pressure range considered the $Fm\bar{3}m$/$R\bar{3}m$ phase remains the most stable. The difference in Gibbs free energy between these two structures is extremely small, meaning synthesis of a pure sample of either could be difficult.

We found the double-delta smearing employed in superconductivity calculations to be of particular importance. Its effect on calculated $T_c$ changes from system to system; in particular, in our calculations the default minimum smearing employed by \textsc{quantum espresso} overestimates $T_c$ for LaH$_{10}$ by $\sim$20\ K and YH$_{10}$ by $\sim$30\ K when compared to optimal smearing.

\section*{Acknowledgements}
We thank Bartomeu Monserrat for helpful discussions. A.M.S.\ acknowledges funding through an EPSRC studentship. M.J.H.\ and M.S.J.\ acknowledge the EPSRC Centre for Doctoral Training in Computational Methods for Materials Science for funding under grant number EP/L015552/1. C.J.P.\ is supported by the Royal Society through a Royal Society Wolfson Research Merit award. R.J.N.\ is supported by EPSRC under Critical Mass Grant EP/P034616/1 and the UKCP consortium grant EP/P022596/1. This work was performed using resources provided by the Cambridge Service for Data Driven Discovery (CSD3) operated by the University of Cambridge Research Computing Service ( \url{www.csd3.cam.ac.uk}), provided by Dell EMC and Intel using Tier-2 funding from the EPSRC (capital grant EP/P020259/1), and DiRAC funding from the STFC (\url{www.dirac.ac.uk}).

\bibliography{references.bib}

\clearpage
\section{Stability and superconductivity of lanthanum and yttrium decahydrides - supplementary material}
\section{Pressure-volume curves}
\begin{figure}
    \centering
    \includegraphics[width=\columnwidth]{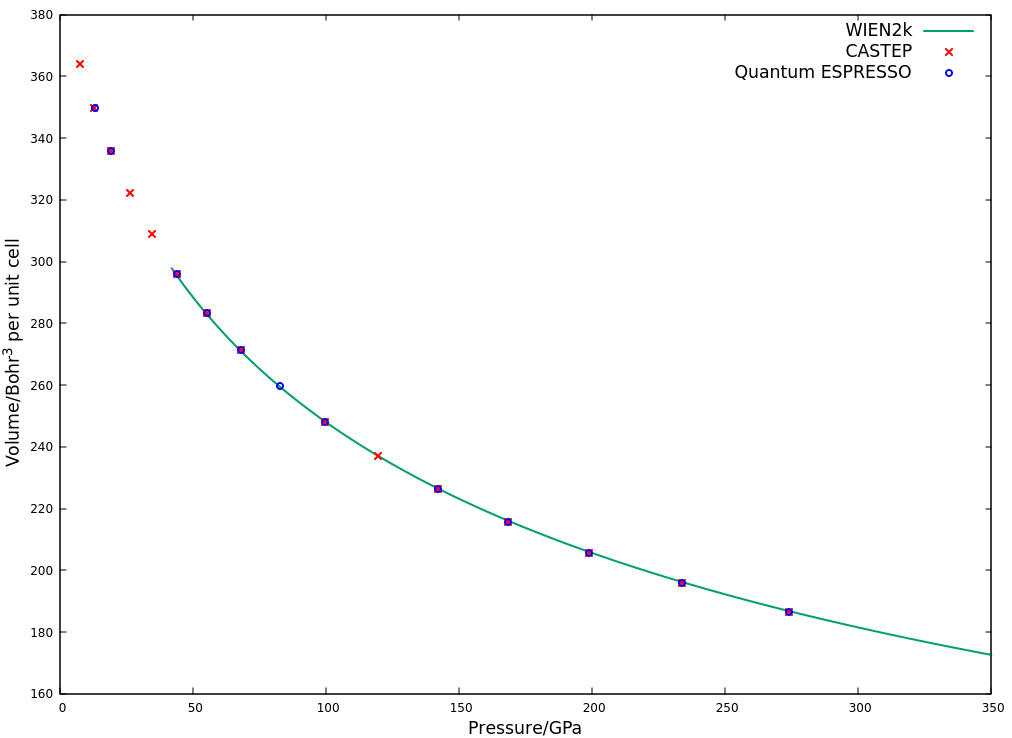}
    \caption{Pressure-volume curve calculated using the all-electron code WIEN2k compared to data obtained using \textsc{castep} and \textsc{quantum espresso} for $Fm\bar{3}m$-LaH$_{10}$. The WIEN2k data was calculated for Ref.\ \cite{liu2018} and was provided to us by Hanyu Liu.}
    \label{fig:lah10_pv}
\end{figure}

\begin{figure}
    \centering
    \includegraphics[width=\columnwidth]{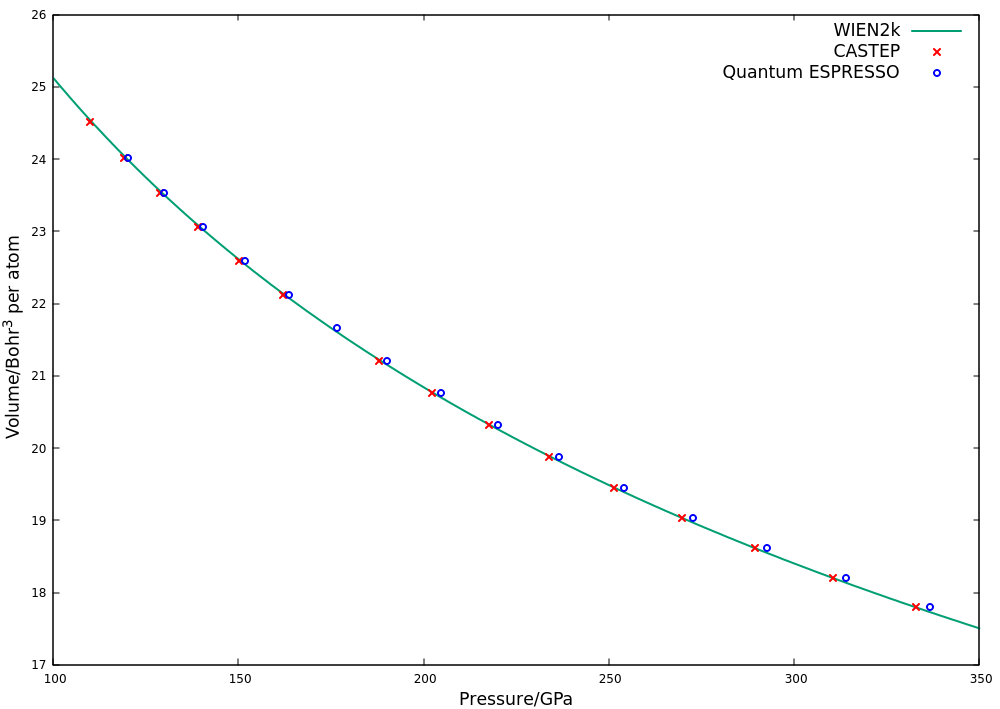}
    \caption{Pressure-volume curve calculated using the all-electron code WIEN2k compared to data obtained using \textsc{castep} and \textsc{quantum espresso} for $Im\bar{3}m$-YH$_6$. The WIEN2k data was calculated for Ref.\ \cite{peng2017} and was provided to us by Feng Peng.}
    \label{fig:yh6_pv}
\end{figure}

In order to validate the pseudopotentials used in \textsc{quantum espresso}, we compared the pressure-volume curves produced for the clathrate structures LaH$_{10}$ and YH$_6$ to data obtained using \textsc{castep} \cite{castep2005} and using the all-electron code WIEN2k \cite{wien2k_code}. Figs.\ \ref{fig:lah10_pv} and \ref{fig:yh6_pv} contain data points produced by the following codes and pseudopotentials:
\begin{itemize}
\item \textsc{quantum espresso} \cite{QE-2009,QE-2017}: Scalar-relativistic, ultrasoft PBE pseudopotentials downloaded from \\ \url{https://www.quantum-espresso.org/pseudopotentials/ps-library}
\item \textsc{castep} \cite{castep2005}: on-the-fly (OTF) pseudopotentials generated with default pseudopotential strings for hydrogen and yttrium and 2$\vert$2.3$\vert$5$\vert$6$\vert$7$\vert$50U:60:51:52:43\{4f0.1\}(qc$=$4.5)[4f0.1] for lanthanum - in terms of generating a pseudopotential capable of exactly reproducing the all-electron PV curve for LaH$_{10}$, we find that the addition of 4f0.1 to the La pseudopotential string is crucial.
\end{itemize}

Good agreement between \textsc{quantum espresso} and WIEN2k is observed in both cases. For the most part, the structure searches in this work used \textsc{castep} QC5 pseudopotentials, instead of OTF pseudopotentials, for computational efficiency. The detailed La-H convex hull at 150\ GPa presented in the main text used \textsc{castep} OTF pseudopotentials generated by the string 2$\vert$2.3$\vert$5$\vert$6$\vert$7$\vert$50U:60:51:52:43\{4f0.1\}(qc$=$4.5)[4f0.1] for La and the default string for H.

\section{Structure searching and convex hulls}
\begin{figure}
    \centering
    \includegraphics[width=\columnwidth]{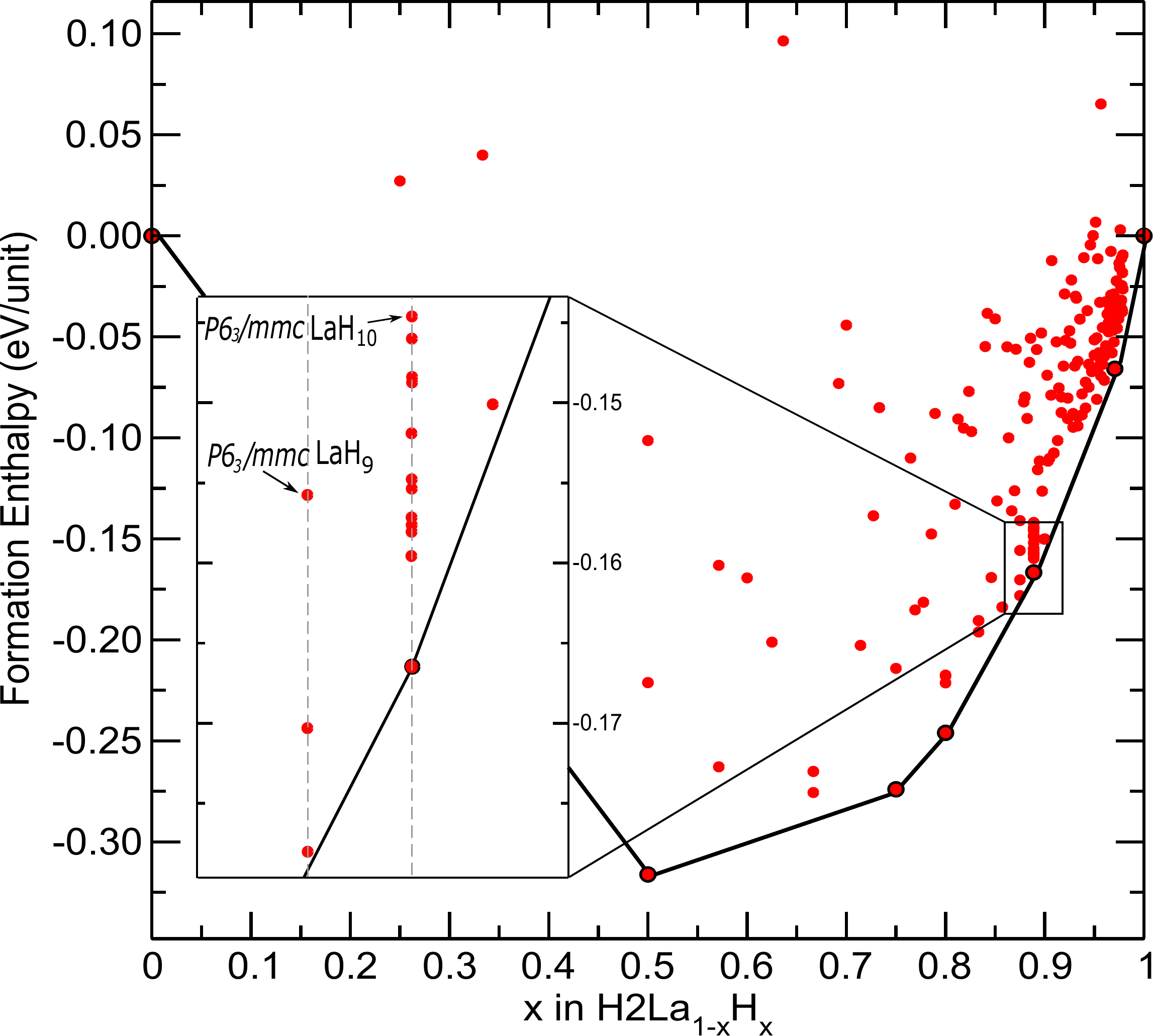}
    \caption{A convex hull for the La-H system at 150\ GPa, calculated with the same parameters as the La-H convex hull in the main text, but without the 0.1 4f electron in the La pseudopotential. Here, LaH$_3$, LaH$_5$, LaH$_6$, LaH$_{10}$ and LaH$_{35}$ are found on the hull, compared to La$_3$H$_{10}$, LaH$_4$, LaH$_5$, LaH$_{11}$ and LaH$_{35}$ in the hull of the main text, highlighting the importance of the 4f electron contribution.}
    \label{fig:lah_hull}
\end{figure}

\begin{figure}
    \centering
    \includegraphics[width=\columnwidth]{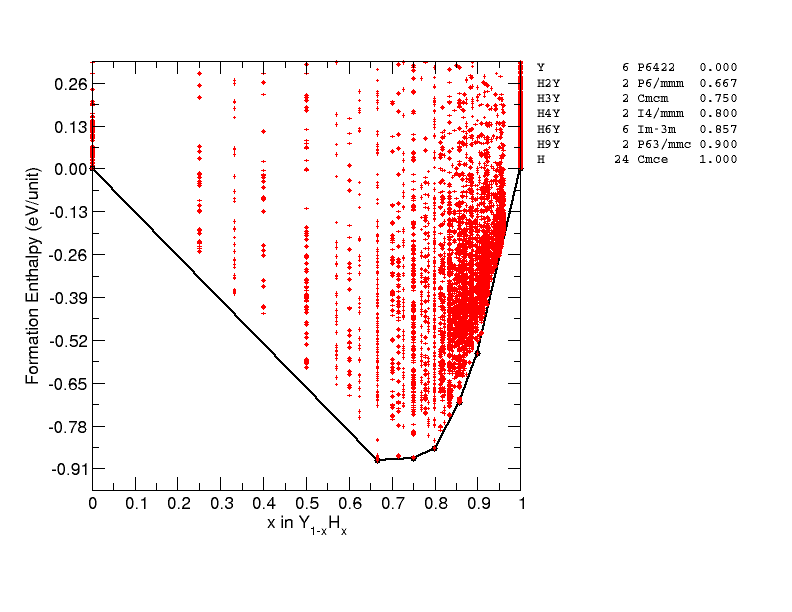}
    \caption{An example convex hull for the Y-H system, showing that YH$_{10}$ is meta-stable at the static lattice level at 400 GPa. We note that our Y-H hull identifies the YH$_4$, YH$_6$ and YH$_9$ structures recently reported in experiment \cite{troyan2019, kong2019} ($I4/mmm$, $Im\bar{3}m$ and $P6_3/mmc$, respectively), highlighting the success of crystal structure prediction methods.}
    \label{fig:yh_hull}
\end{figure}

\begin{figure}
    \centering
    \includegraphics[width=\columnwidth]{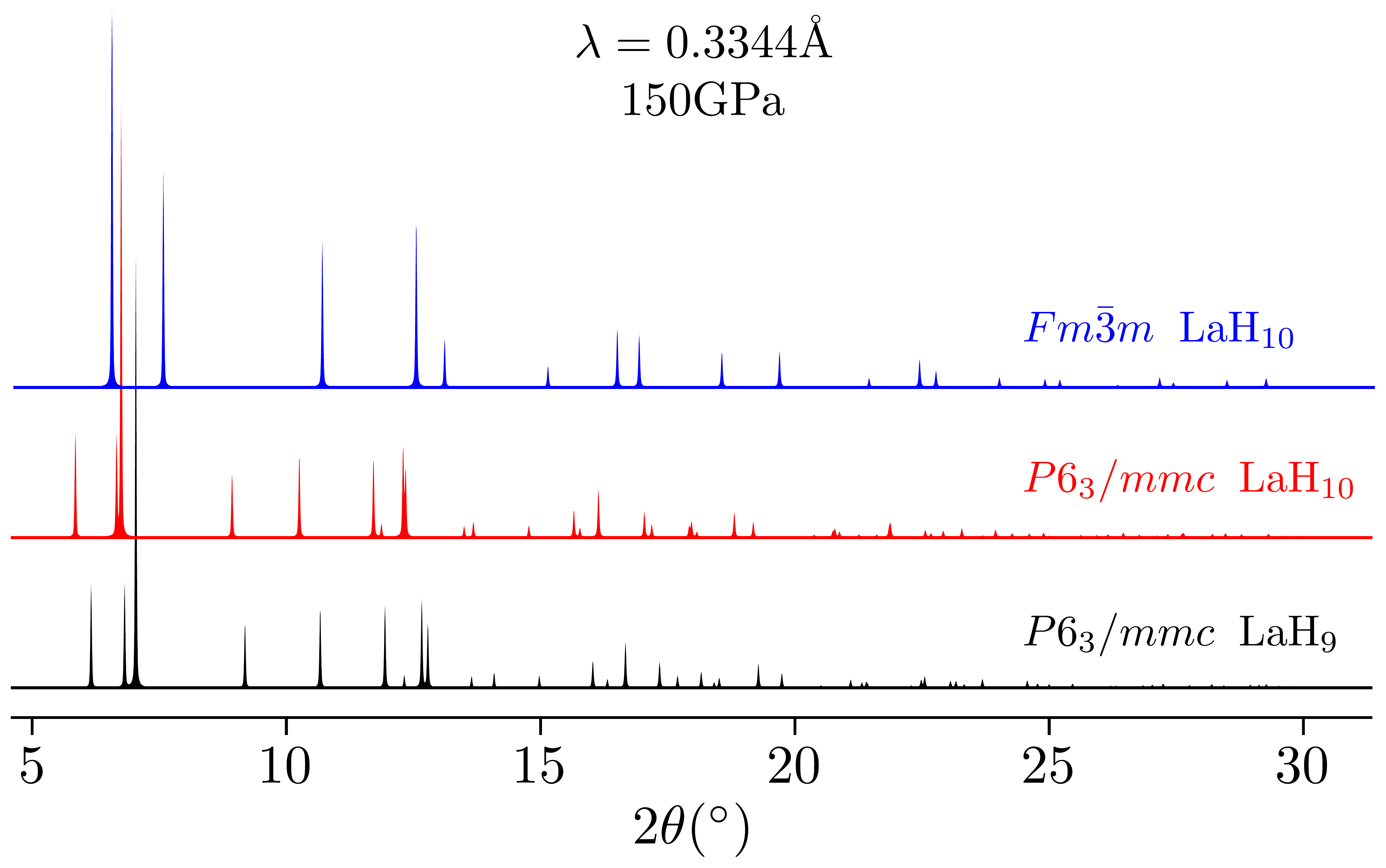}
    \caption{Simulated X-ray powder diffraction patterns for three of the structures discussed in this work.}
    \label{fig:powder_patterns}
\end{figure}

We constructed well-converged convex hulls for the La-H and Y-H systems using AIRSS \cite{pickard2011, needs2016} and qhull \cite{barber1996}, as illustrated in Figs.\ \ref{fig:lah_hull} and \ref{fig:yh_hull}. Our convex hulls confirm the findings of previous work showing that the LaH$_{10}$ and YH$_{10}$ stoichiometries are on or close to the hull over the pressure ranges we study here ($\sim$150-500 GPa for LaH$_{10}$ and $>\sim$300 GPa for YH$_{10}$) \cite{peng2017}. Powder X-ray diffraction patterns for structures relating to hexagonal impurities are shown in Fig.\ \ref{fig:powder_patterns}. At 150\ GPa, we find a c/a ratio of 1.526 for $P6_3/mmc$-LaH$_{10}$ and of 1.564 for $P6_3/mmc$-LaH$_9$. At the same pressure, the volume per formula unit for $P6_3/mmc$-LaH$_{10}$ is 33.15 \r{A}$^3$ and for $P6_3/mmc$-LaH$_9$ is 31.73 \r{A}$^3$.

Throughout this work, we found the \textsc{c2x} software \cite{c2x} extremely useful for converting between \textsc{castep} and \textsc{quantum espresso} file formats and reporting symmetries at various tolerances.

\section{DFT energies}
\begin{figure}
    \centering
    \includegraphics[width=\columnwidth]{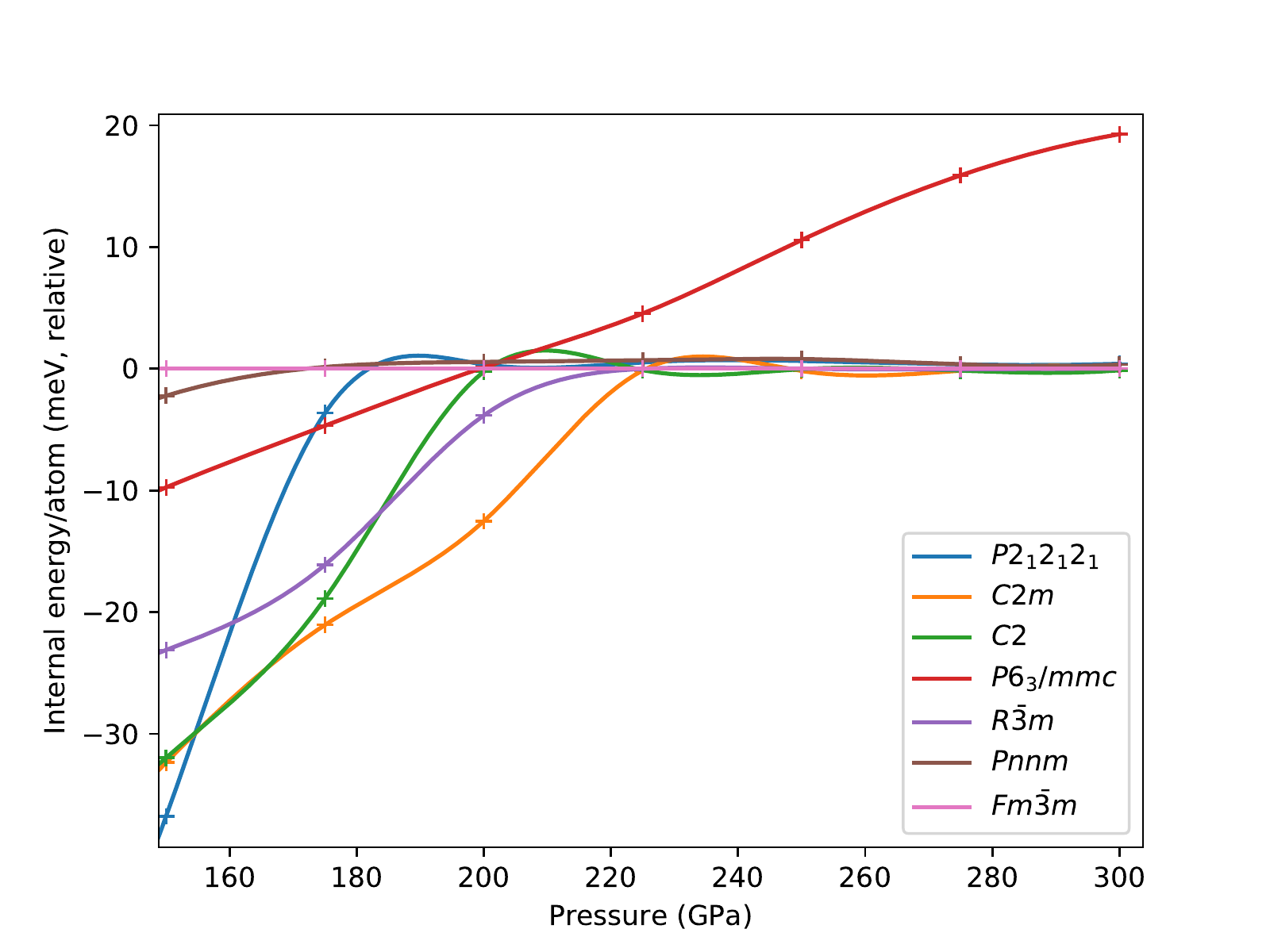}
    \caption{DFT internal energies of LaH$_{10}$ phases, neglecting phonon contributions. We see that at lower pressures, several distortions of the $Fm\bar{3}m$ phase have lower internal energies - we therefore predict that within the harmonic approximation these distorted phases have a range of stability below $\sim$210\ GPa when $Fm\bar{3}m$ becomes dynamically unstable. The electronic energy shows that these distortions approach $Fm\bar{3}m$ at higher pressures. This behaviour is reflected in the Gibbs free energy plot in the main text and can be observed by considering symmetry tolerances between the structures.}
    \label{fig:lah10_dft_energies}
\end{figure}

\begin{figure}
    \centering
    \includegraphics[width=\columnwidth]{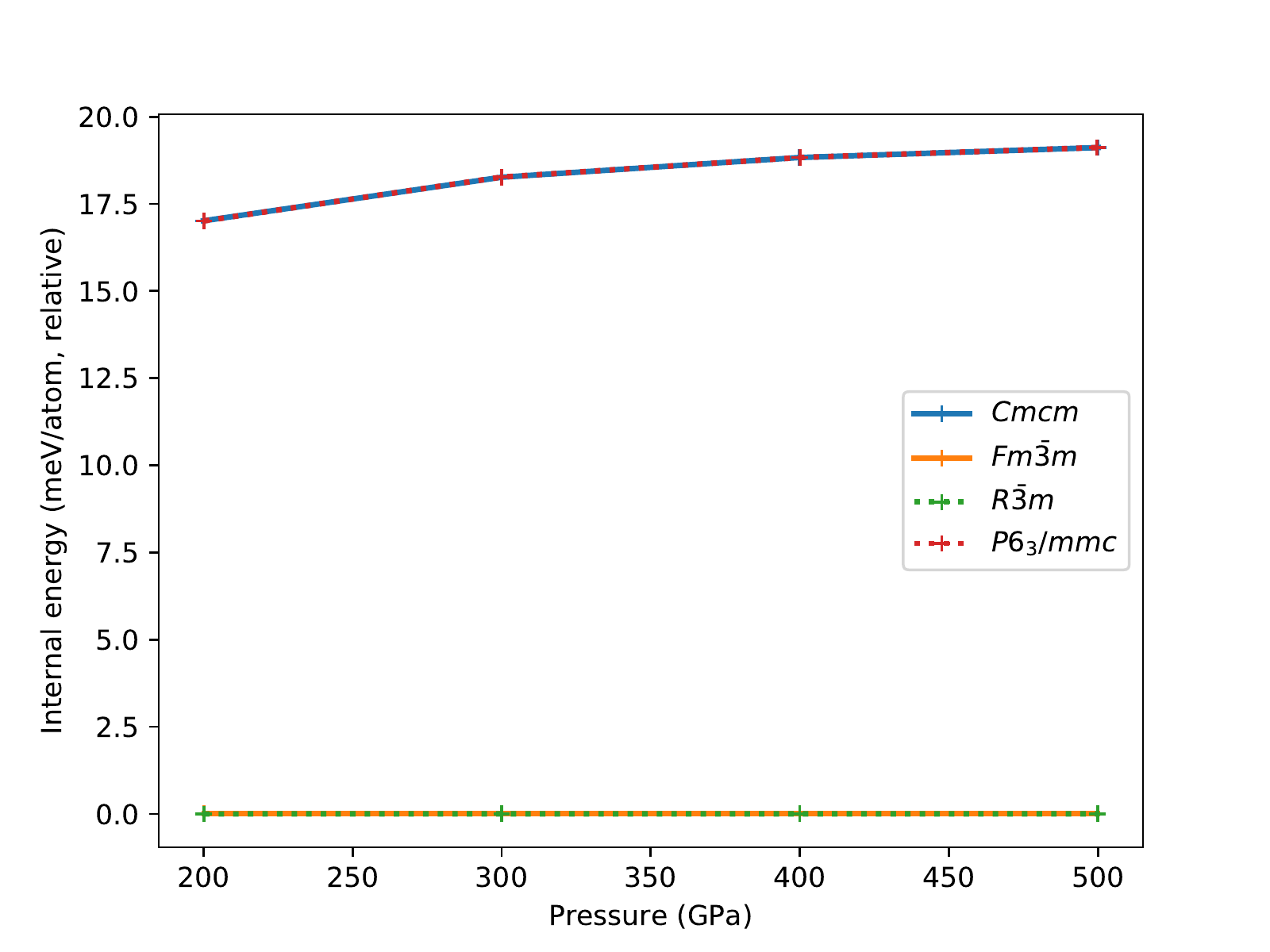}
    \includegraphics[width=\columnwidth]{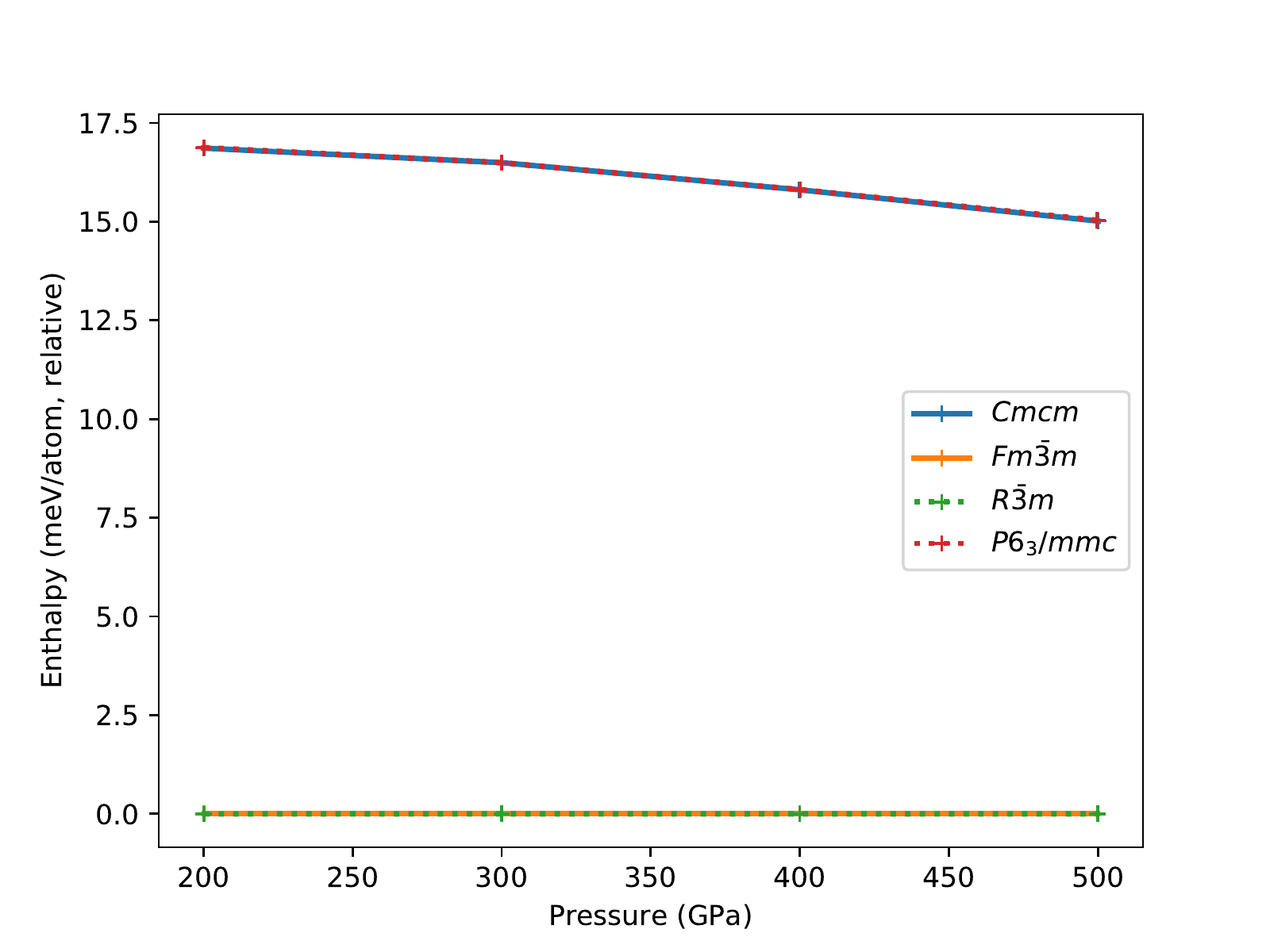}    
    \caption{DFT energetics of YH$_{10}$ phases, neglecting phonon contributions. Top panel: internal energy. Bottom panel: enthalpy. By considering symmetry tolerances between the structures of interest and studying the internal energy, it is clear that the $Fm\bar{3}m$ and $R\bar{3}m$ structures are very similar, as are the $P6_3/mmc$ and $Cmcm$ structures. The energy difference between the two sets of structures at the static-lattice level is too large to be compensated for by differences in phonon contributions to the energy and we do not see any region of stability for $P6_3/mmc$ or $Cmcm$ over the pressure range of interest, as can be seen in the Gibbs free energy plot in the main text.}
    \label{fig:yh10_dft_energies}
\end{figure}

Fig.\ \ref{fig:lah10_dft_energies} shows the relative energies of the LaH$_{10}$ phases, neglecting phonon contributions. Fig.\ \ref{fig:yh10_dft_energies} shows the same for YH$_{10}$ phases. The structure files for all structures studied in this work are available at \url{https://doi.org/10.17863/CAM.46481}.

\begin{figure}
    \centering
    \includegraphics[width=\columnwidth]{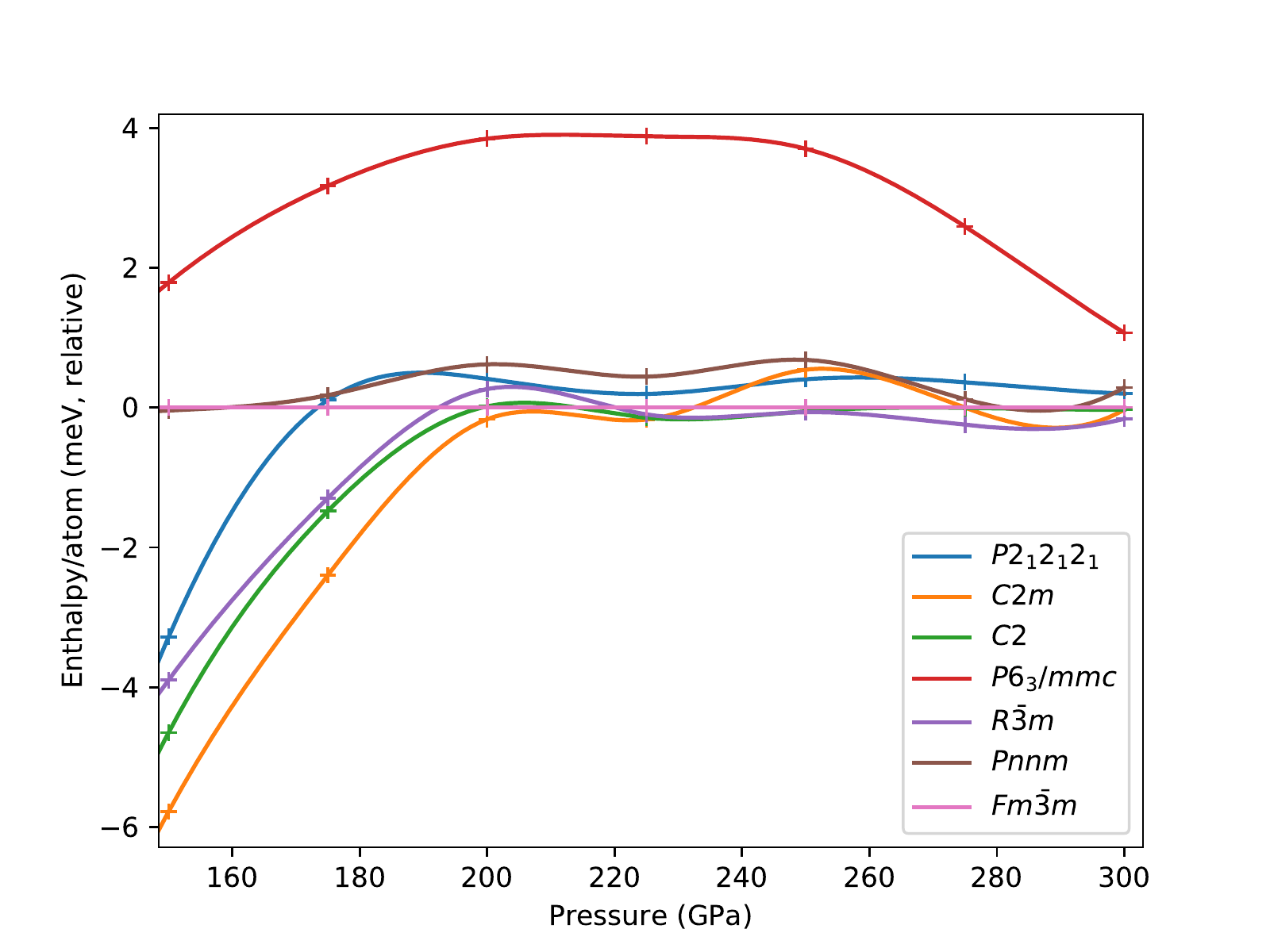}
    \caption{DFT enthalpies of LaH$_{10}$ phases, including the large-unit-cell $Pnnm$, $P2_12_12_1$ and $C2$ phases. We see that these large-cell phases are energetically competitive.}
    \label{fig:lah10_large_cell_enthalpies}
\end{figure}

\section{Convergence testing}
\begin{figure*}
    \centering
    \includegraphics[width=1.7\columnwidth]{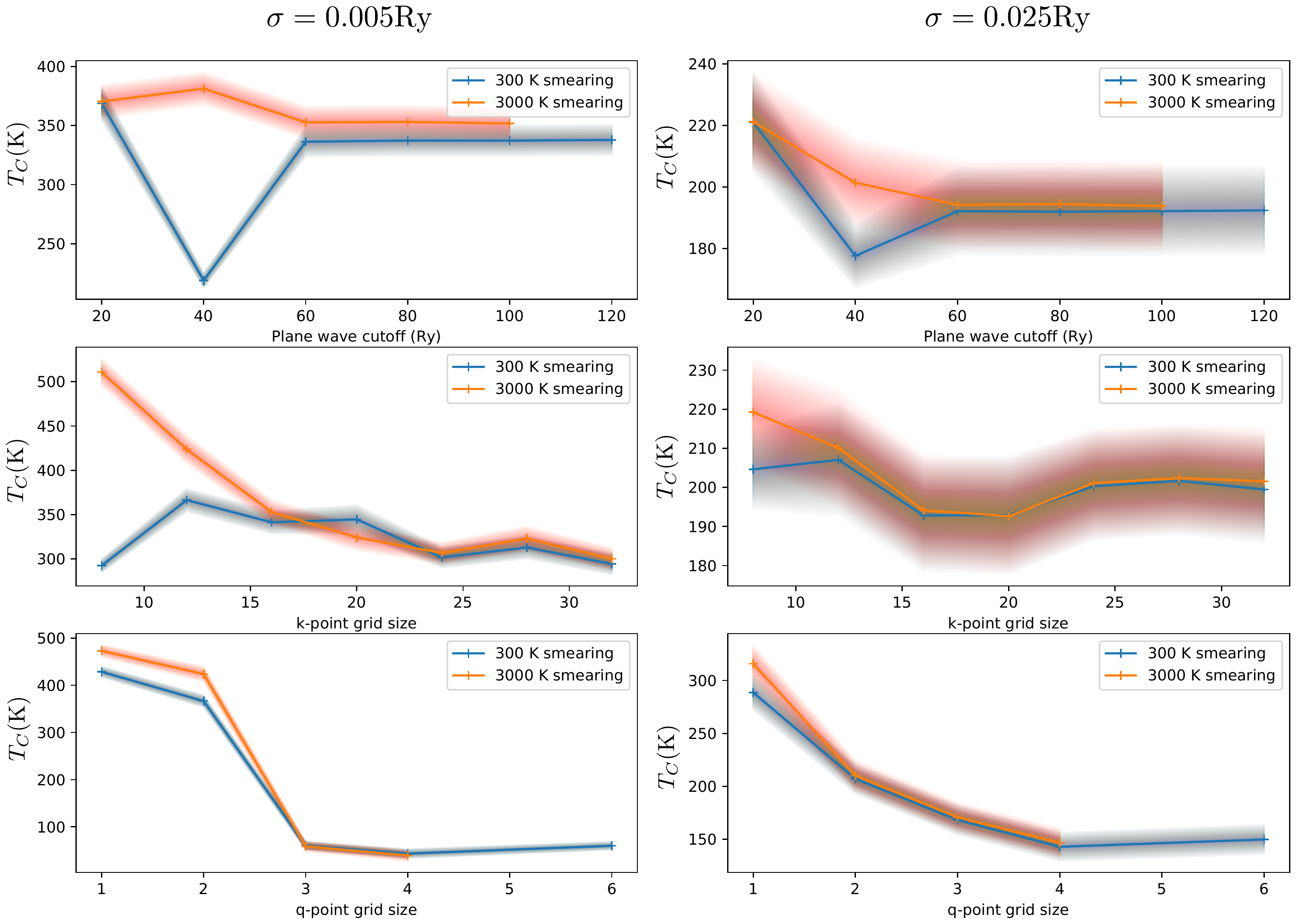}
    \caption{Convergence properties for the critical temperature of the $Fm\bar{3}m$ phase of LaH$_{10}$ for different values of double-delta smearing $\sigma$. The blurred section around each line represents the distribution of $T_C$ for different values of the $\mu^*$ parameter.}
    \label{fig:fm3m_tc_conv}
\end{figure*}

\begin{figure}
    \centering
    \includegraphics[width=\columnwidth]{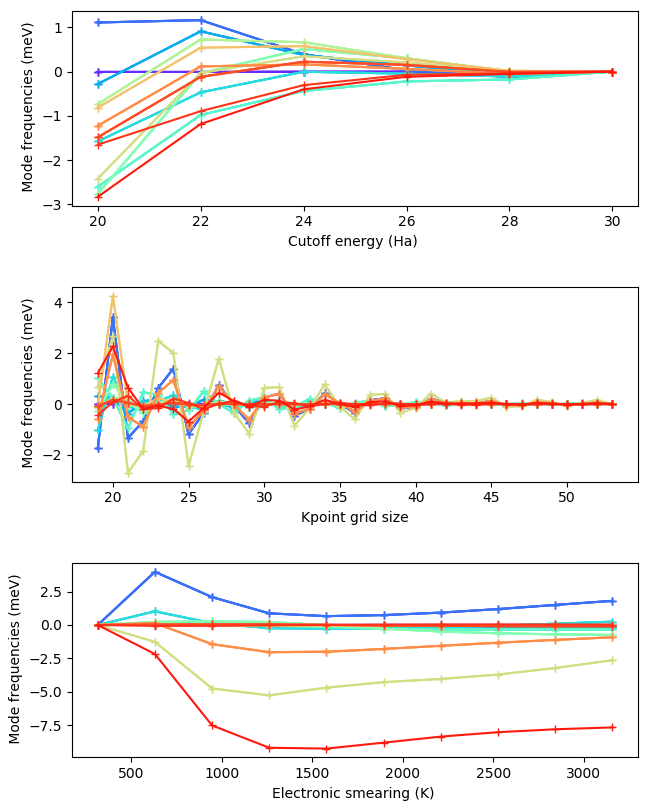}
    \caption{Convergence behaviour of phonon frequencies for $Fm\bar{3}m$ LaH$_{10}$. Each line corresponds to a different phonon mode.}
    \label{fig:phonon_mode_freq_conv}
\end{figure}

\begin{figure}
    \centering
    \includegraphics[width=\columnwidth]{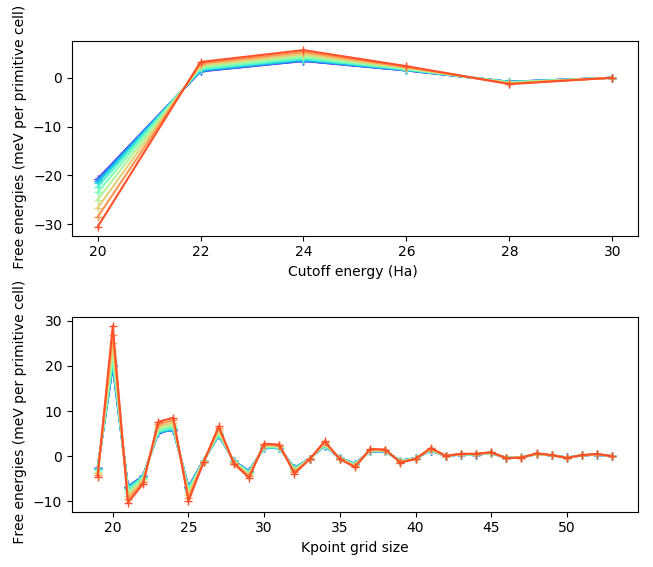}
    \caption{Convergence behaviour of Helmholtz free energies (including vibrational contributions) for $Fm\bar{3}m$ LaH$_{10}$. Different lines correspond to different temperatures.}
    \label{fig:free_energy_conv}
\end{figure}

The first stage towards calculating accurate phase diagrams and superconducting critical temperatures is to establish the computational parameters required to achieve the desired accuracy. In this work, the most relevant parameters were
\begin{itemize}
    \item Electronic plane-wave cutoff
    \item Electronic $\mathbf{k}$-point sampling density
    \item Electronic occupation smearing width/scheme
    \item Phonon $\mathbf{q}$-point sampling density
\end{itemize}
The electronic occupation smearing width was found to be of particular importance. It is important to note that this is the Fermi-surface smearing used during SCF convergence, not the smearing used to evaluate double-delta integrals in superconductivity theory. For typical DFT \textit{energy} calculations the smearing of the electronic Fermi surface often makes a negligible difference, even with high effective temperatures. This is because the energy scale of the electronic band structure is typically equivalent to thousands of Kelvin, due to Fermi statistics pushing electrons into higher and higher energy states \cite{steve_textbook}. As the total energies are only sensitive to the average change in the occupied energy states, and because the smearing is symmetric around the Fermi surface, we can use high smearing temperatures when we are only interested in total energies. Unfortunately the same is not true for electron-phonon coupling parameters. Only states close to the Fermi surface contribute significantly to electron-phonon coupling. Therefore, in order to calculate accurate electron-phonon coupling properties, we need an accurate resolution of the (unsmeared) Fermi surface. To describe a Fermi surface accurately, we therefore require good Brillouin zone resolution, which can be achieved using large $\mathbf{k}$-point grids. This can be seen in Fig.\ \ref{fig:fm3m_tc_conv}. The resulting parameter set for LaH$_{10}$ is a 60 Ry cutoff, a $\mathbf{k}$-point grid with a spacing of 2$\pi\times$0.015 \r{A}$^{-1}$ (equivalent to a $24\times24\times24$ grid for the $Fm\bar{3}m$ phase) and a $\mathbf{q}$-point grid that is 8 times smaller than the k-point grid (equivalent to a $3\times3\times3$ grid for the $Fm\bar{3}m$ phase). From Fig.\ \ref{fig:fm3m_tc_conv}, we see that the difference between using 300\ K and 3000\ K smearing leads to a greater error in $T_c$ than using an approximate $\mu^*$, but does not noticeably impact convergence; this allows us to use 300\ K smearing without significant loss of efficiency. In order to carry out these calculations within a reasonable time-frame, we have optimised the electron-phonon coupling code in \textsc{quantum espresso}, leading to a $10\times$ speedup for our systems. These changes have been submitted (and accepted) to the \textsc{quantum espresso} project to allow others to benefit from our modifications. Since the electron-phonon calculations require such high convergence parameters, thermodynamic quantities are already well-converged with the chosen parameters, as shown in Fig.\ \ref{fig:phonon_mode_freq_conv} and \ref{fig:free_energy_conv}.

\section{Treatment of double-delta smearing}
\begin{figure}
    \centering
    \includegraphics[width=\columnwidth]{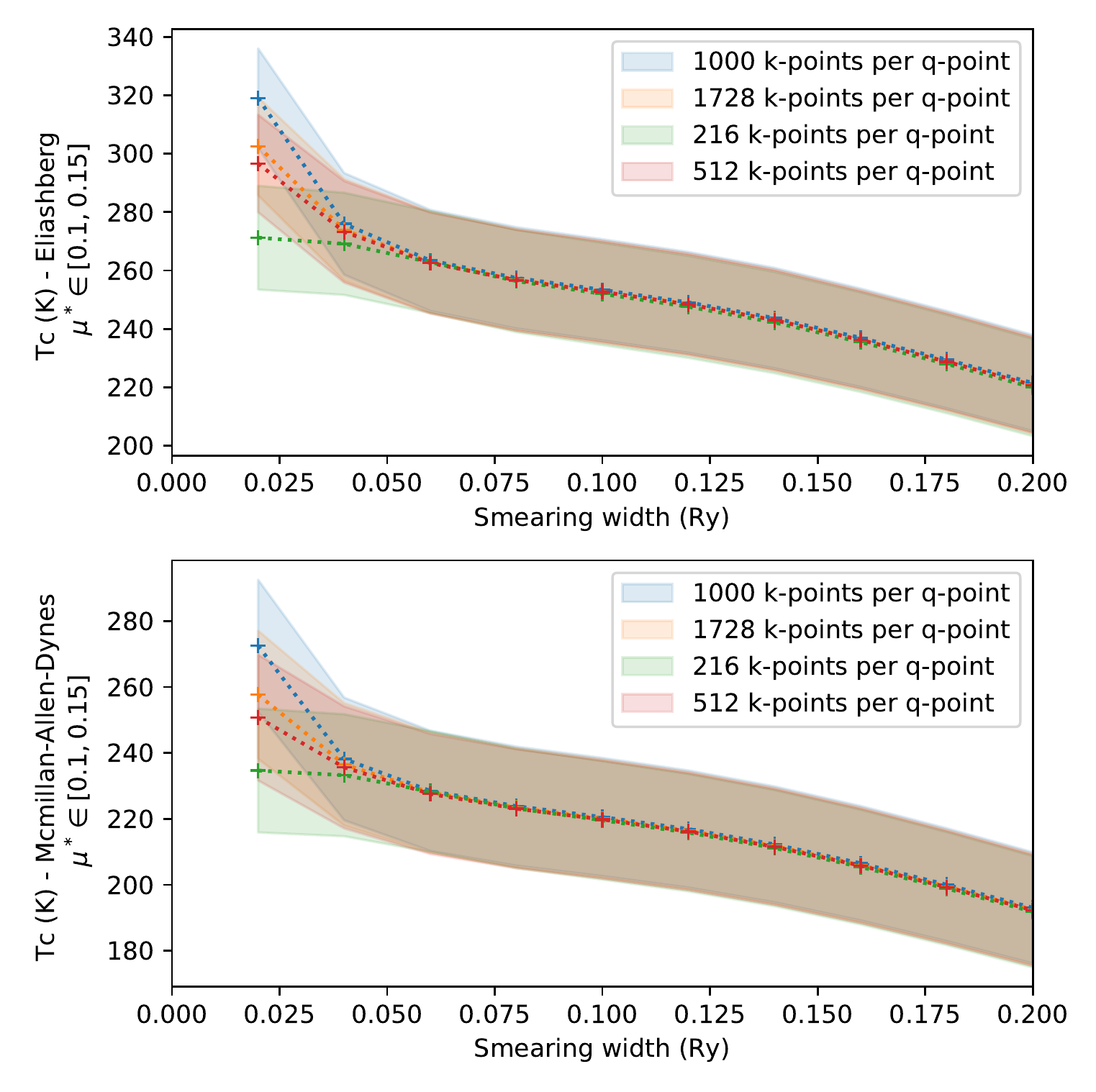}
    \caption{$T_C$ vs. double-delta smearing for $Fm\bar{3}m$ YH$_{10}$, calculated using different sized k-point grids with a fixed $3\times3\times3$ q-point grid.}
    \label{fig:kpts_per_qpt_fm3m_yh10}
\end{figure}

As noted in the main text, we use a multiple-grid scheme to ensure that the double-delta smearing parameters used are appropriate. It is straightforward to see that below a certain smearing value strong discrepancies between different grid sizes arise; see Fig.\ \ref{fig:kpts_per_qpt_fm3m_yh10} for example. We have modified our version of \textsc{quantum espresso} so that we can increase the number of double-delta smearing values used, which ensures that we can always identify this region of insufficient smearing.

\section{Phonon dispersion curves}
\begin{figure*}
    \centering
    \includegraphics[width=\textwidth]{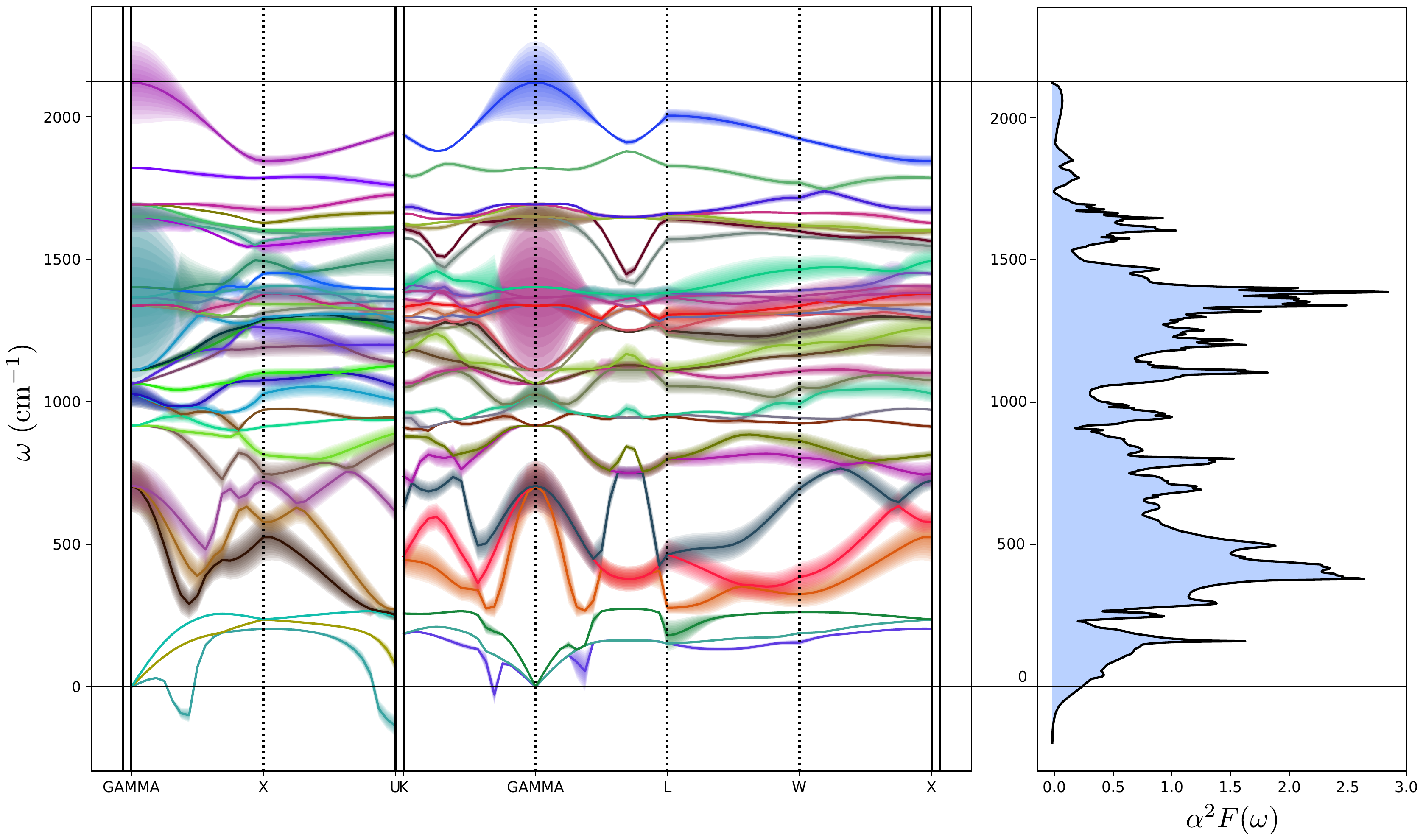}
    \caption{The phonon dispersion, showing linewidth broadening and the Eliashberg function for the $Fm\bar{3}m$ phase of LaH$_{10}$ at 200 GPa.}
    \label{fig:phonon_linewidths}
\end{figure*}

Fig.\ \ref{fig:phonon_linewidths} shows the phonon dispersion, linewidths and resulting Eliashberg function for the $Fm\bar{3}m$ phase of LaH$_{10}$ at 200 GPa. This is the highest pressure at which imaginary phonon modes are present, and where we apply our procedure for estimating $T_C$ in the presence of imaginary modes.

\begin{figure}
    \centering
    \includegraphics[width=\columnwidth]{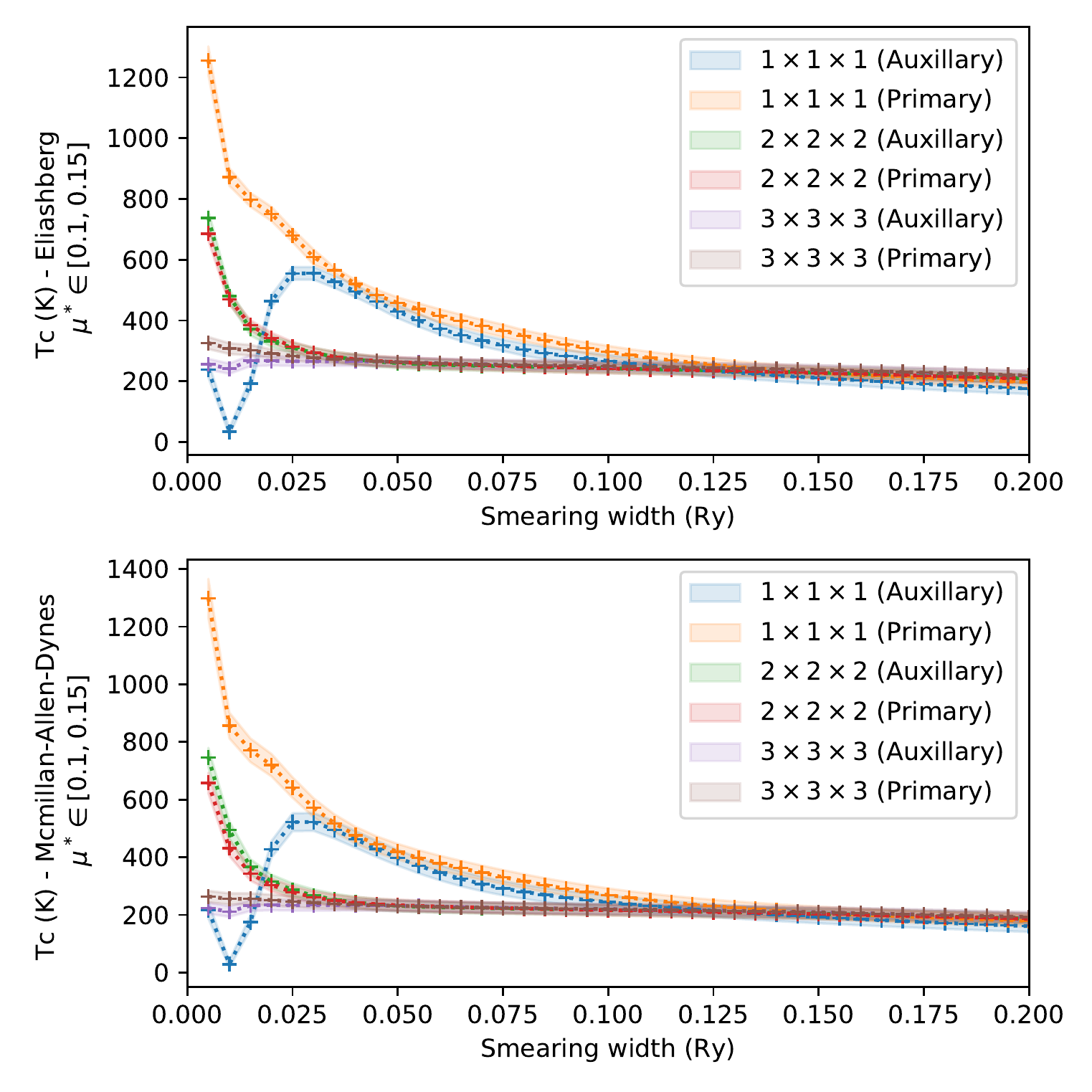}
    \caption{$T_C$ vs. double-delta smearing for $Fm\bar{3}m$ YH$_{10}$, calculated using different sized q-point grids. The results for primary/auxiliary k-point grids for each q-point grid are shown (primary = 512 k-points per q-point, auxiliary = 261 k-points per q-point).}
    \label{fig:qpt_convergence_fm3m_yh10}
\end{figure}

\begin{figure}
    \centering
    \includegraphics[width=\columnwidth]{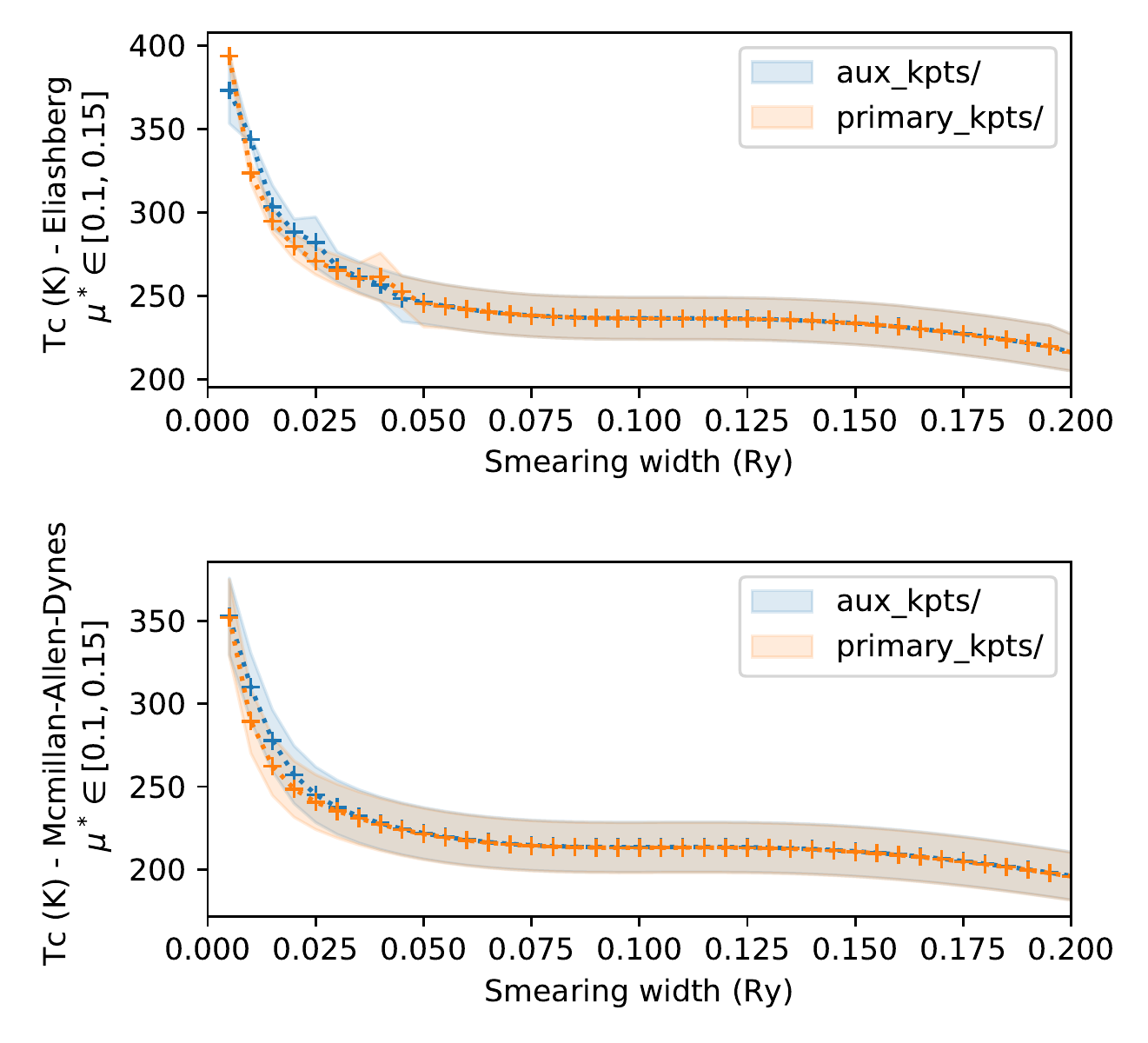}
    \caption{$T_C$ vs. double-delta smearing for $Fm\bar{3}m$ LaH$_{10}$, calculated using different sized k-point grids with a fixed $3\times3\times3$ q-point grid.}
    \label{fig:kpts_per_qpt_fm3m_lah10}
\end{figure}

\section{Background theory: Electron-Phonon coupling in DFT}
Typically, within DFT the nuclear coordinates, $R$, are treated as fixed and the electronic Kohn-Sham system is solved within the fixed nuclear potential. In order to calculate the effects of electron-phonon coupling within the DFT formalism we must consider leading-order corrections to the Born-Oppenheimer approximation in nuclear displacements. Expanding our Kohn-Sham potential in terms of these displacements leads to
\begin{equation}
\label{eq:ks_pertubation}
    V_{KS}(R+\delta R) = V_{KS}(R) + \sum_{\kappa, p} \frac{\partial V_{KS}}{\partial R_{\kappa, p}} \cdot \delta R_{\kappa, p} + O(\delta R^2).
\end{equation}
where $R_{\kappa, p}$ is the position of atom $\kappa$ in unit cell $p$. An atomic displacement of an atom can be written in terms of phonon creation and annihilation operators \cite{DFT_ELEC_PHONON} as
\begin{equation}
    \delta R_{\kappa, p} = \frac{1}{\sqrt{N_pM_\kappa}} \sum_{q\nu} e^{iq\cdot R_p} \frac{1}{\sqrt{2\omega_{q\nu}}} \left( a_{q\nu}+a_{-q\nu}^\dagger \right) e_{\kappa\nu}(q)
\end{equation}
where $e_{k\nu}(q)$ and $\omega_{q,\nu}$ are, respectively, the eigenvector and frequency of the phonon mode with creation operator $a_{q\nu}^\dagger$. $R_p$ is the position of the $p$\textsuperscript{th} unit cell within the periodic cell, of which there are $N_p$. $M_\kappa$ is the mass of atom $\kappa$. Substituting this into Eq.\ \ref{eq:ks_pertubation} we obtain
\begin{equation}
    V_{KS}(R + \delta R) = V_{KS}(R) + \frac{1}{\sqrt{N_p}} \sum_{q\nu} G_{q\nu} (a_{q\nu} + a_{-q\nu}^\dagger)
\end{equation}
where
\begin{equation}
    G_{q\nu} = \frac{1}{\sqrt{2\omega_{q\nu}}} \sum_\kappa \frac{e_{\kappa\nu}(q)}{\sqrt{M_\kappa}} \cdot \sum_p e^{iq \cdot R_p} \frac{\partial V_{KS}}{\partial R_{\kappa,p}}
\end{equation}
This allows us to write down the resulting electron-phonon coupling Hamiltonian in second-quantized form as 
\begin{equation}
\begin{aligned}
    &H_{ep} (\delta R) \\&= \sum_{nkn'k'} \bra{n,k} V_{KS}(R + \delta R) - V_{KS}(R) \ket{n', k'} c_{nk}^\dagger c_{n'k'} \\
    &= \sum_{q\nu} \left[\sum_{nkn'k'} \bra{n,k} G_{q\nu} \ket{n'k'} c_{nk}^\dagger c_{n',k'}\right] \frac{a_{q\nu} + a_{-q\nu}^\dagger}{\sqrt{N_p}}
\end{aligned}
\end{equation}
where $c_{nk}^\dagger$ creates a Kohn-Sham electron in orbital $n$, wavevector $k$ (i.e., occupies the Bloch state $u_{nk}(x)\exp(ik\cdot x)/\sqrt{N_p}$). Substituting our definition of $G_{q\nu}$ we have
\begin{equation}
\label{eq:big_g_matrix_elements}
\begin{aligned}
    &\bra{n,k} G_{q\nu} \ket{n'k'} \\&= \frac{1}{\sqrt{2\omega_{q\nu}}} \sum_\kappa \frac{e_{\kappa\nu}(q)}{\sqrt{M_\kappa}} \cdot \sum_p e^{iq \cdot R_p} \bra{n,k} \frac{\partial V_{KS}}{\partial R_{\kappa,p}} \ket{n',k'}
\end{aligned}
\end{equation}
Now
\begin{equation}
\begin{aligned}
    &\bra{n,k} \frac{\partial V_{KS}}{\partial R_{\kappa,p}} \ket{n',k'} \\&= \frac{1}{N_p} \int u_{nk}^*(x) e^{-ik\cdot x} \frac{\partial V_{KS}}{\partial R_{\kappa,p}}(x) u_{n'k'}(x)e^{ik'\cdot x} \;dx \\
    &=\frac{1}{N_p} \int u_{nk}^*(x-R_p)e^{-ik\cdot (x-R_p)} \frac{\partial V_{KS}}{\partial R_{\kappa,p}}(x-R_p) \\&\hspace{2cm} \times u_{n'k'}(x-R_p)e^{ik'\cdot (x-R_p)} \;dx \\
    &= e^{iR_p\cdot(k-k')} \int_{\text{1\textsuperscript{st} unit-cell}} \hspace{-1cm} u_{nk}^*(x)e^{-ik\cdot x} \frac{\partial V_{KS}}{\partial R_{\kappa,0}}(x) u_{n'k'}(x)e^{ik'\cdot x} \;dx
\end{aligned}
\end{equation}
where in the last line we have used Bloch's theorem and the fact that
\begin{equation}
    \frac{\partial V_{KS}}{\partial R_{\kappa,p}}(x - R_p) = \frac{\partial V_{KS}}{\partial R_{\kappa,0}}(x)
\end{equation}
where $R_{\kappa,0}$ is the position of atom $\kappa$ in the first unit cell. We may now write Eq.\ \ref{eq:big_g_matrix_elements} as
\begin{equation}
\begin{aligned}
    &\bra{n,k} G_{q\nu} \ket{n'k'} \\&= \frac{1}{\sqrt{2\omega_{q\nu}}} \sum_\kappa \frac{e_{\kappa\nu}(q)}{\sqrt{M_\kappa}} \cdot \bra{n,k} \frac{\partial V_{KS}}{\partial R_{\kappa,0}} \ket{n',k'}_{\text{uc}} \\ &\hspace{2cm}\times \underbrace{\sum_p e^{i(q + (k-k')) \cdot R_p}}_{N_p\delta_{q,k-k'}}
\end{aligned}
\end{equation}
where the subscript ``uc" on the ket means integration only over the first unit cell. Finally we obtain the DFT electron-phonon coupling Hamiltonian
\begin{equation}
\begin{aligned}
    H_{ep} = \frac{1}{\sqrt{N_p}} \sum_{q\nu knm} \bra{m,k+q} G_{q\nu,\text{uc}} \ket{n,k}_{\text{uc}} \\ \times c_{m,k+q}^\dagger c_{n,k} (a_{q\nu} + a_{-q\nu}^\dagger)
\end{aligned}
\end{equation}
where we have defined
\begin{equation}
    G_{q\nu,\text{uc}} = \frac{1}{\sqrt{2\omega_{q\nu}}} \sum_\kappa \frac{e_{\kappa\nu}(q)}{\sqrt{M_\kappa}} \cdot \frac{\partial V_{KS}}{\partial R_{\kappa,0}}
\end{equation}
This allows us to write down the Hamiltonian for an interacting Kohn-Sham-electron-phonon system, correct to first order in electron-phonon coupling constants $g_{mn\nu}(k,q) = \bra{m,k+q} G_{q\nu,\text{uc}} \ket{n,k}_{\text{uc}}$:
\begin{equation}
\label{eq:electron_phonon_hamiltonian}
\begin{aligned}
    H &= \underbrace{\sum_{kn} \epsilon_{nk} c_{nk}^\dagger c_{nk}}_{\text{Electronic dispersion}} + \underbrace{\sum_{q\nu} \omega_{q\nu} \left(a_{q\nu}^\dagger a_{q\nu} + \frac{1}{2}\right)}_{\text{phonon dispersion}} +\\
    &\underbrace{\frac{1}{\sqrt{N_p}} \sum_{kqmn\nu} g_{mn\nu}(k,q)c_{m,k+q}^\dagger c_{nk} \left( a_{q\nu} + a_{-q\nu}^\dagger\right).}_{\text{electron-phonon coupling}}
\end{aligned}
\end{equation}
From the parameters in this Hamiltonian we can also define the electron-phonon coupling strength associated with each phonon mode, $\lambda_{q\nu}$, and the isotropic Eliashberg spectral function, $\alpha^2F(\omega)$
\begin{equation}
\begin{aligned}
    \lambda_{q,\nu} &= \frac{1}{N(\epsilon_F)\omega_{q\nu}\Omega_{\text{BZ}}} \\&\times \sum_{nm} \int_{\text{BZ}} |g_{mn\nu}(k,q)|^2 \delta(\epsilon_{n,k} - \epsilon_F)\delta(\epsilon_{m,k+q} - \epsilon_F) dk
\end{aligned}
\end{equation}
\begin{equation}
    \alpha^2F(\omega) = \frac{1}{2\Omega_{\text{BZ}}}\sum_\nu \int_{\text{BZ}} \omega_{q\nu}\lambda_{q\nu}\delta(\omega - \omega_{q\nu}) dq
\end{equation}
from which we may calculate the critical temperature by solution of the Eliashberg equations \cite{eliashberg1960}. The only additional requirement is the Morel-Anderson pseudopotential \cite{mu_star}, which we treat as an empirical parameter with values between 0.1 and 0.15.

\clearpage
\bibliography{references.bib}
\end{document}